\documentclass[twocolumn]{aastex631}

\newcommand{\foo}{\makebox[0pt]{\textbullet}\hskip-0.5pt\vrule width 1pt\hspace{\labelsep}}

\usepackage{stackengine}

\usepackage{amsmath}
\usepackage{gensymb}
\usepackage{makecell}
\renewcommand{\tablecomments}[1]{\vspace{5pt} \noindent {\footnotesize #1}}

\graphicspath{{./}{figures/}}

\begin{document}

\title{Ultraviolet Imaging of SR 12 c with HST/WFC3:  

Accretion and Variability of a Giant Planet at the End Stages of Growth} 

\correspondingauthor{Claire O. Finley}
\email{clairefinley@ucsb.edu} 

\author[0000-0002-9884-9584]{Claire O. Finley} \affiliation{Department of Astronomy, University of Texas at Austin, 2515 Speedway C1400, Austin, TX 78712, USA} \affiliation{Department of Physics, University of California, Santa Barbara, Santa Barbara, CA 93106, USA} 

\author[0000-0003-2649-2288]{Brendan P. Bowler} \affiliation{Department of Physics, University of California, Santa Barbara, Santa Barbara, CA 93106, USA} \affiliation{Department of Astronomy, University of Texas at Austin, 2515 Speedway C1400, Austin, TX 78712, USA}

\author[0000-0002-4392-1446]{Ya-Lin Wu} \affiliation{Department of Physics, National Taiwan Normal University, Taipei 116, Taiwan}

\author[0000-0001-9811-568X]{Adam L. Kraus} \affiliation{Department of Astronomy, University of Texas at Austin, 2515 Speedway C1400, Austin, TX 78712, USA}

\author[0000-0003-2969-6040]{Yifan Zhou} \affiliation{University of Virginia, 530 McCormick Road, Charlottesville, VA 22904, USA}

\author[0000-0003-0568-9225]{Yuhiko Aoyama} \affiliation{School of Physics and Astronomy, Sun Yat-sen University, Zhuhai 519082, People's Republic of China}

\author[0000-0003-0562-1511]{William Best} \affiliation{Department of Astronomy, University of Texas at Austin, 2515 Speedway C1400, Austin, TX 78712, USA}

\author[0000-0002-1483-8811]{Ian Czekala} \affiliation{School of Physics \& Astronomy, University of St. Andrews, North Haugh, St. Andrews KY16 9SS, UK}

\author[0000-0001-9227-5949]{Catherine C. Espaillat} \affiliation{Department of Astronomy, Boston University, 725 Commonwealth Avenue, Boston, MA 02215, USA}

\author[0000-0002-7821-0695]{Katherine B. Follette} \affiliation{Department of Physics \& Astronomy, Amherst College, 25 East Drive, Amherst, MA 01002, USA}

\author[0000-0002-7154-6065]{Gregory J. Herczeg} \affiliation{Kavli Institute for Astronomy and Astrophysics, Peking University, Beijing 100871, People’s Republic of China}\affiliation{Department of Astronomy, Peking University, Beijing 100871, People’s Republic of China}

\author[0000-0001-6301-896X]{Raquel A. Martinez} \affiliation{Department of Physics and Biophysics, University of San Diego, San Diego, CA 92110, USA}

\author[0000-0003-1639-510X]{Connor E. Robinson} \affiliation{Department of Astronomy, Boston University, 725 Commonwealth Avenue, Boston, MA 02215, USA}

\author[0000-0001-6532-6755]{Quang H. Tran} \affiliation{Department of Astronomy, Yale University, New Haven, CT 06511, USA}

\author[0000-0002-4479-8291]{Kimberly Ward-Duong} \affiliation{Department of Astronomy, Smith College, Northampton, MA 01063 USA}

\begin{abstract} 
Many details of the gas accretion phase during giant planet formation remain untested. We present new 0.2--0.7~$\mu$m UV-through-red optical imaging of the young, wide-orbit planetary-mass companion SR 12 c from the Wide Field Camera 3 (WFC3) instrument on board the Hubble Space Telescope. SR 12 c exhibits strong accretion-related continuum excess blueward of $\sim$5000~\AA\ and clear signs of the Balmer jump at 3646~\AA. We derive a total accretion luminosity of 1.65~$\pm$~$0.19 \times 10^{-5} L_{\odot}$ and a mass accretion rate of 8~$\pm$~$2\times 10^{-12}$~M$_{\odot}$~yr$^{-1}$. Based on its mass and age, SR 12 c will not grow by an appreciable amount at its current accretion rate; it is at the end stages of assembly. No accretion variability is evident between the two epochs of the WFC3 observations spanning a month-long baseline, but the H$\alpha$ emission line strength decreases by 90\% compared to the reported flux from five years earlier. Combined with previous observations of SR 12 c, we assemble one of the most complete spectral energy distributions of a young giant planet to date, spanning the UV through sub-mm wavelengths (0.2--880 $\mu$m). This adds SR 12 c to the small yet growing sample of planets with detailed accretion and disk constraints, which together are beginning to establish the diversity of timescales and physical processes governing the formation of giant planets. 
\end{abstract}


\section{Introduction} \label{sec:intro}
Gas giants are the most readily detectable subset of planets, but details about their origin and evolution have been challenging to address with observations. In particular, the exact timing of giant planet formation and the mechanism through which mass is accreted via the circumplanetary disk onto the growing planet remains elusive \citep{Benisty2023, Adams2025b}. 

There are two primary proposed pathways to form giant planets: core accretion and disk instability. The core/pebble accretion model \citep{Pollack1996, Inaba2003, Alibert2005, Lambrechts2012, Adams2025a} posits that giant planets form from the ``bottom up" through $\sim$14 orders of magnitude in size and $\sim$40 orders of magnitude in mass. When the core reaches a critical mass (predicted to be $\sim$5-20 M$_{\oplus}$; \citealt{Papaloizou1999, Rafikov2006}), hydrostatic equilibrium is disrupted, and the planet rapidly accretes gas from the circumstellar disk to form a giant planet. However, this method struggles to explain the presence of giant planets on very wide orbits of hundreds to thousands of AU beyond even the largest protoplanetary disks without invoking dynamical ejection (e.g., \citealt{Bowler2016}).

In the alternative ``top down" model of disk instability, disks that are sufficiently massive and cool are expected to fragment into low-mass self-gravitating objects \citep{Boss1997}. If these protoplanets continue to accrete, they may grow to become brown dwarfs or low-mass stars. Those that do not substantially gain mass can be left in the planetary regime, representing the tail end of the stellar and substellar companion mass function \citep{Kratter2010}. Simulations of these fragmented clumps suggest that substellar objects created via disk instability may have more massive disks and higher rates of accretion than those created by the direct collapse of isolated gas cores \citep{StamHerc2015}, presenting an observational test that links accretion and the formation mechanism of giant planets \citep{Durisen2007}. 

The bulk of a giant planet's accreted mass is funneled through its circumplanetary disk (CPD; \citealt{Keith2014, Szulagyi2017, Maeda2022}). There is growing evidence that CPDs are intrinsically common: \cite{Bowler2017} found that 46 $\pm$ 14\% of $<$20 M$_\mathrm{Jup}$ companions with ages $<$15 Myr show Pa$\beta$ emission, a sign of active accretion, while \cite{Martinez2022} found a disk fraction of 56 $\pm$ 12\% for a comparable range of mass and age. The best way to understand mass transfer onto a growing giant planet is by studying accreting protoplanets at early ages ($\leq$10 Myr, \citealt{Follette2023}). The majority of a planet's accretion luminosity ($L_\mathrm{acc}$) lies in the ultraviolet continuum if the shocked gas is optically thick \citep{Valenti1993, Calvet1998, Zhu2015}. If the shock is optically thin, hydrogen line emission can contribute significantly to the total accretion luminosity \citep{Aoyama2020}. To capture the total $L_\mathrm{acc}$, observations from space provide the broadest wavelength coverage into the blue-optical and UV (e.g., \citealt{Zhou2014}) with instruments such as the Wide Field Camera 3 (WFC3) on the Hubble Space Telescope (HST). The $\sim$2000--5000 \AA\ range covering the Balmer jump at 3646~\AA, offers a convenient way to constrain the UV excess and provides a more complete picture of accretion compared to accretion-related emission lines (such as H$\alpha$ or Pa$\beta$, e.g. \citealt{Bowler2011, Demars2023, Zhou2025, Close2025a, Close2025b}). This tactic has been explored extensively in the stellar and brown dwarf regimes (e.g., \citealt{RomanDuval2020, Wendeborn2024}), but similar observations are rare for young giant planets \citep{Zhou2014, Zhou2021}.

At longer wavelengths, the mid-infrared (IR) region is sensitive to reprocessed radiation from CPDs around growing giant planets. CPDs are understudied observationally because relatively few have been detected at thermal and millimeter wavelengths, suggesting CPDs are more optically thick or have more compact radii than expected \citep{Bowler2015, Wu2015, Wolff2017, Isella2019, Wu2020, Benisty2021, Andrews2021, Wu2022, Scholz2023, Patapis2025, Cugno2025, Hoch2025}. In particular, it is not clear when or how these disks disperse, the diversity of their physical properties, and whether grain growth and moon formation are common phenomena \citep{Benisty2021}. 

The most thoroughly-studied protoplanets reside in the PDS 70 system, which hosts PDS 70 b \citep{Keppler2018, Muller2018} and c \citep{Haffert2019}, two accreting protoplanets \citep{Wagner2018, Zhou2021} embedded in the transition disk of this $\sim$5 Myr Sun-like host star. \cite{Isella2019} tentatively detected a compact mm source at the position of PDS 70 c, which was recovered in follow-up ALMA observations by \cite{Benisty2021} at high significance, providing the most compelling evidence for a CPD surrounding the planet. \cite{Blakely2025} explored the mid-IR excess emission of PDS 70 c with JWST/NIRISS and found that models that include a CPD offer a good fit to its 1--5 $\mu$m SED.\footnote{Recently, \cite{DominguezJamett2025} suggested that the sub-mm emission from PDS 70 c may instead correspond to free-free emission from a jet rather than thermal emission from a CPD based on multi-wavelength ALMA observations.} More recently, the discoveries of 2MASS J16120668-301027 b \citep{Li2025} and WISPIT 2b \citep{vanCapelleveen2025, Close2025b} have increased the sample of young accreting planets to study planet-disk interactions and directly witness young planets gaining mass. Although these systems have provided an unprecedented window into the planet formation process, their high planet-to-star contrasts and close angular separations have made it challenging to carry out detailed studies of the accretion and disk properties of these planets.  

At wider orbital distances, direct imaging campaigns have identified a population of planetary-mass companions (PMCs) at hundreds to thousands of AU from their host stars \citep{Bowler2016}. The census of wide planets now amounts to over two dozen gas giants ($\sim$10-20 $M_\mathrm{Jup}$). These distant planets complicate the traditional view of planet formation because they could represent planets that formed closer in and were subsequently ejected through dynamical encounters \citep{Veras2009, Bowler2011, Pearce2019}, or they may be the tail end of direct collapse from a filament or molecular cloud core \citep{Bate2009, Offner2023, Xuan2024}. Irrespective of formation route, at these separations it is easier to disentangle planetary accretion signatures and thermal excess emission from the host star, making these distant planets ideally situated to jointly study accretion and CPDs (e.g., \citealt{Bowler2011, Bailey2013, Lachapelle2015}).  

\subsection{The Young Wide Companion SR 12 c}
At a mean distance\footnote{No astrometric solution is presented for SR 12 AB in Gaia DR3, likely because it is a binary. A parallax is reported in Gaia DR2, but the single-star solution is likely unreliable. As such, we adopt the mean cluster distance.} of $139 \pm 5$ pc \citep{Ratzenboeck2023}, the $\rho$ Ophiuchi cloud complex is a low-density star-forming region that contains several hundred young stars, including the close (0$\farcs$2) binary SR 12 AB (2MASS J16271951–2441403) in the Lynds 1688 (L1688) cloud \citep{Wilking2008}. \cite{Bowler2014} used evolutionary models \citep{Baraffe1998} to estimate masses of $\approx$1.0 and $\approx$0.5 $M_{\odot}$ for SR 12 A and B, respectively, based on the age of $\sim$2 Myr from \cite{Kuzuhara2011}, component luminosities from \cite{Wahhaj2010}, and component effective temperatures from \cite{LuhmanRieke1999}. The low-mass companion to SR 12 AB, SR 12 c, was discovered and confirmed to have common proper motion with its host binary by \cite{Kuzuhara2011}. Reported values for its model-inferred mass have varied based on the estimated distance, luminosity, and age. Here, we adopt 16 $\pm$ 2 M$_\mathrm{Jup}$ from Wu et al. (submitted) which assumes a slightly older system age of 3--4 Myr.\footnote{The estimated age of low-extinction sources toward the L1688 cloud in $\rho$ Oph (such as SR 12) has been updated to be slightly older at $\sim$3-4 Myr \citep{EsplinLuhman2020}.} SR 12 c is on an extremely wide orbit, separated from its host binary by a projected separation of $\approx$1200 AU (8$\farcs$7, \citealt{Kuzuhara2011}). This makes it readily accessible to detailed follow-up observations to characterize the planet and its circumstellar environment (see Section~\ref{sec:obsdat}).\footnote{Note that the host binary does not show evidence of a protoplanetary disk given its lack of significant excess in \textit{Spitzer} photometry beyond 24 $\mu$m \citep{Wahhaj2010} and its 3$\sigma$ dust-mass limit of $<$0.007 M$_{\oplus}$ in ALMA Band 7 \citep{Wu2022}. Moreover, any remaining protoplanetary disk gas accreting onto the companion or replenishing its disk is unlikely at these wide separations.} 

Following its discovery, an abundance of imaging and spectroscopic observations targeting SR 12 c have been carried out. \cite{Kuzuhara2011} presented \textit{J-, H-,} and \textit{K$_s$-}band photometry from SIRIUS, the near-IR camera mounted on the InfraRed Survey Facility in Sutherland, South Africa. This was supplemented with \textit{H-, K-} and \textit{L'-} band photometry from Subaru's Infrared Camera and Spectrograph (IRCS) and Coronagraphic Imager with Adaptive Optics (CIAO), which confirmed the red colors, low temperature, and planetary mass of the companion---assuming hot-start evolutionary models in which the forming planet preserves much of its initial entropy \citep{Marley2007}.

Follow-up IRTF/SpeX near-IR spectra of SR 12 c from \cite{Bowler2014} revealed an angular \textit{H}-band shape--a sign of low surface gravity and youth---and a spectral type of M9 $\pm$ 0.5. They also note an absence of substantial Pa$\beta$ emission at 1.28 $\mu$m, ruling out the type of strong accretion evident in several other wide planetary-mass companions (e.g., \citealt{Bowler2011, Zhou2014, Wu2015, Betti2022}).

Using the X-shooter instrument on the Very Large Telescope (VLT), \cite{SantamariaMiranda2018} presented a 0.3 to 2.5 $\mu$m spectrum for SR 12 c, which revealed prominent hydrogen emission lines, including exceptionally strong H$\alpha$ emission signifying ongoing accretion, with an estimated total accretion rate of $\approx$10$^{-11}$ M$_{\odot}$ yr$^{-1}$ \citep{SantamariaMiranda2019}. They also carried out a maximum likelihood fit of BT-Settl models from \cite{Allard2014} and determine a log $g$ of 4.0 $\pm$ 0.5 and an effective temperature of 2600 $\pm$ 100 K.

Probing further into the infrared, \cite{Martinez2022} presented Spitzer/IRAC photometry, which confirmed the mid-IR excess observed in SR 12 c, indicating the presence of a warm CPD surrounding the planet. The CPD dust was subsequently detected with the Atacama Large Millimeter Array (ALMA) by \cite{Wu2022} at 880 $\mu$m, corresponding to $\approx$1 lunar mass of dust, assuming a mean disk temperature of 24 K and optically thin emission. The estimated dust mass is about two times greater than the PDS 70 c disk \citep{Benisty2021}. 

In a companion paper (Wu et al., submitted), we present new complementary 5.6--21 $\mu$m imaging from the Mid-Infrared Instrument (MIRI) on board the James Webb Space Telescope (JWST). This mid-IR photometry is best described by a two-component blackbody model, although irradiated flat disk and viscous accretion models also provide a good match to the data. We also find a lack of strong silicate emission at 10 $\mu$m suggesting significant grain growth in the CPD, and a mid-IR SED pointing to a later disk evolutionary stage.

Here, we present new UV-through-red optical imaging of SR 12 c with HST/WFC3. This extends the SED of SR 12 c to ultraviolet wavelengths, making it the broadest-sampled SED for a young giant planet to date. In Section~\ref{sec:obsdat}, we describe the observations in 5 filters spanning $\sim$2200 to 6700 \AA, as well as an updated estimate of the interstellar extinction to the SR 12 system using photometry of the host binary. Section~\ref{sec:mods} summarizes properties of the model grids used to fit the accretion and photosphere components of SR 12 c. Section~\ref{sec:disc} discusses the inferred accretion rate and variability of SR 12 c. We conclude in Section~\ref{sec:conc} with a summary of our results.  

\begingroup 
\nolinenumbers
\begin{figure}
    \centering
    \includegraphics[width =0.9\columnwidth]{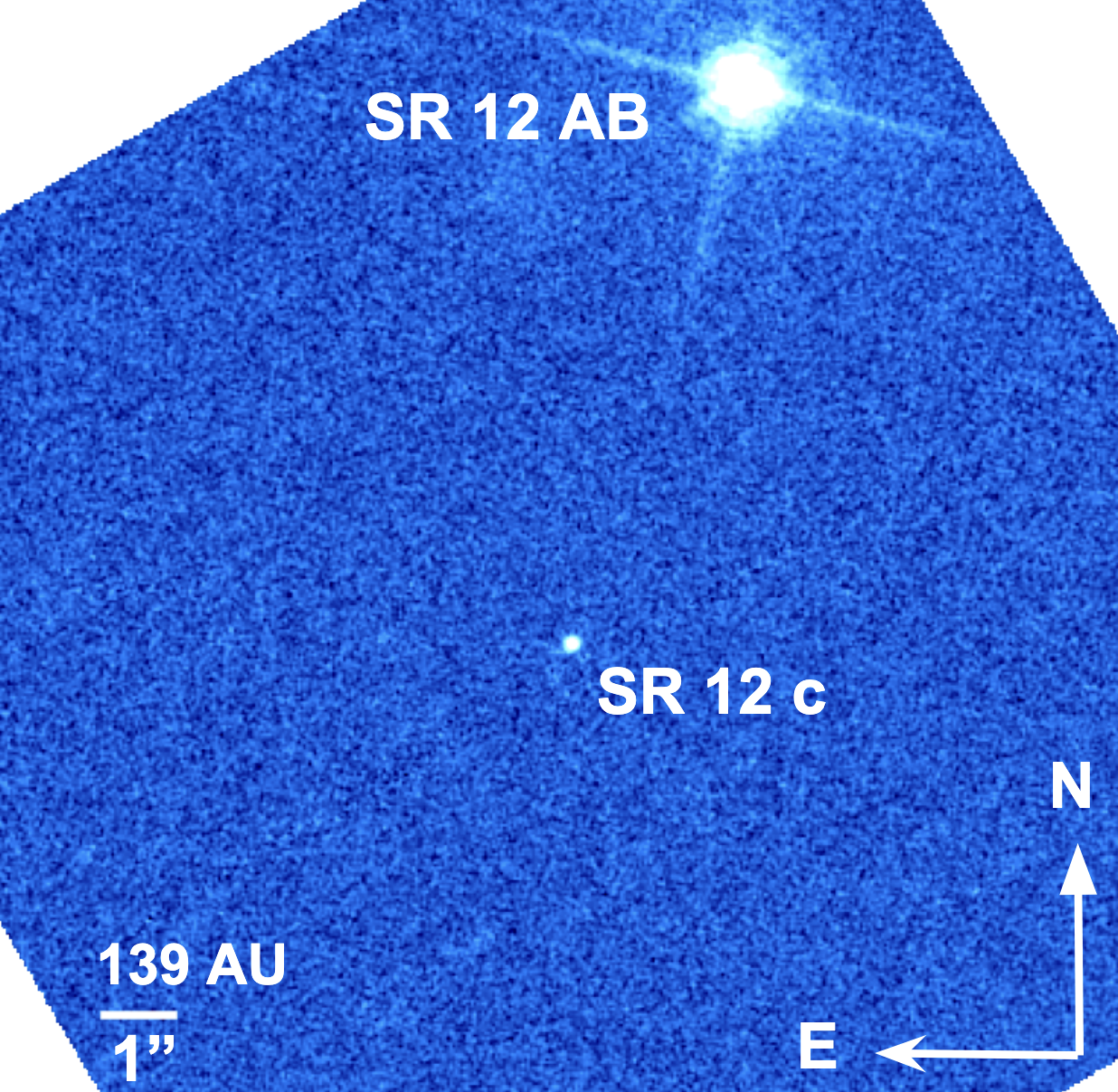}
    \caption{SR 12 c and its host binary, SR 12 AB in the H$\alpha$ filter (F656N) from HST/WFC3. North is up and East is left.} 
    \label{fig:img}
\end{figure}
\endgroup

\section{Observations and Data Analysis}\label{sec:obsdat}

\subsection{HST WFC3/UVIS Imaging}\label{sec:hstuv}
SR 12 c was imaged in 5 filters from $\sim$2200-6700 \AA\ with the UVIS arm of the WFC3 instrument on board the Hubble Space Telescope as a part of the general observer (GO) program 16302 (PI: Wu) on two dates: UT 2021 February 28 and UT 2021 March 26. A three-point line dither pattern was used with a point spacing of 0$\farcs$135. The observations utilized the SUB-C512C subarray of the UVIS2 channel, a 513 x 512 subarray near amplifier C. SR 12 c was observed in the F225W and F336W filters to sample wavelengths shorter than the Balmer break (see Section~\ref{sec:slab} below), the F438W and F555W filters to measure the optical continuum redward of the Balmer break, and the F656N filter to measure flux from the H$\alpha$ emission line. Integration times and signal-to-noise ratios for the companion are listed in Table~\ref{tab:hstobs}. An image of the system architecture is shown in Figure~\ref{fig:img}. Images in each filter are shown for the host binary and the companion in Figure~\ref{fig:tile}. 

\begingroup 
\nolinenumbers
\begin{figure*}
    \centering
    \includegraphics[width =\textwidth]{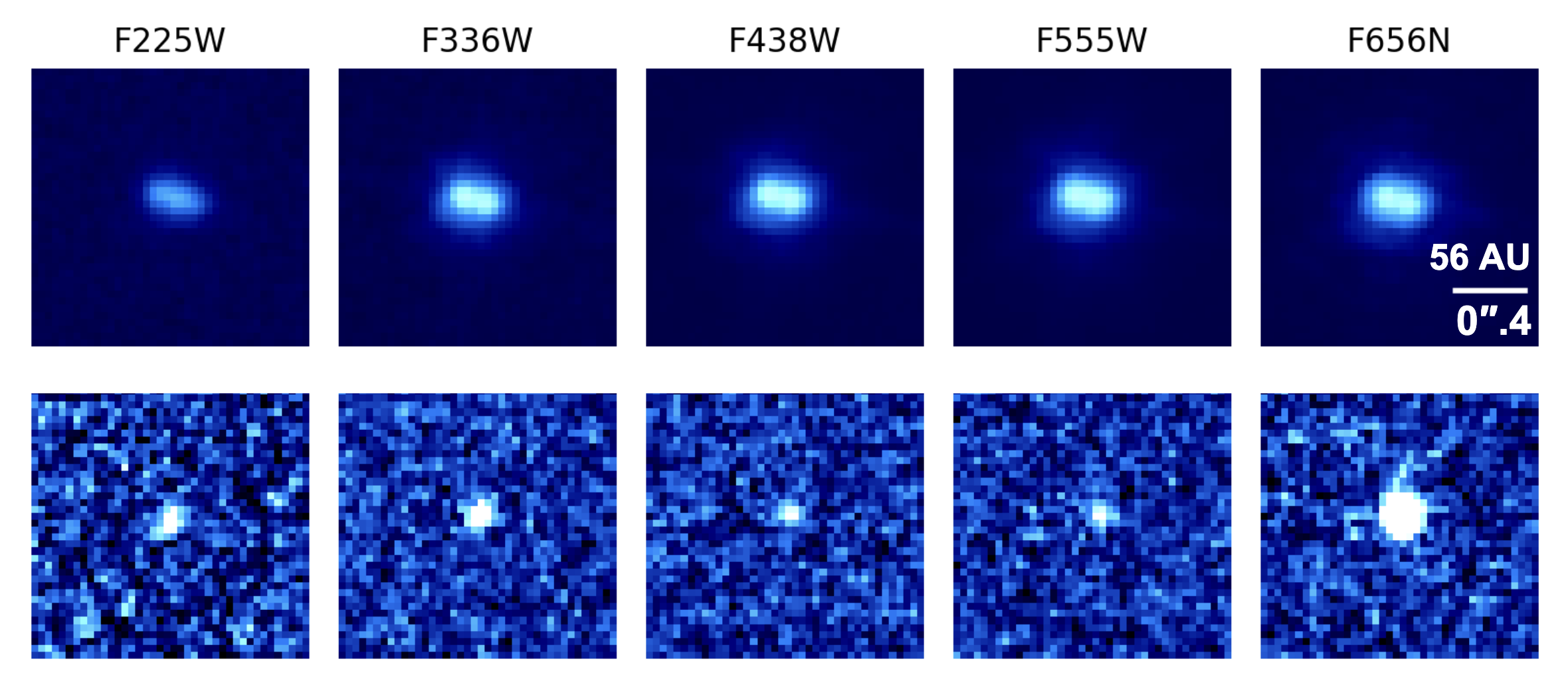}
    \caption{Top: HST/WFC3-UVIS images of SR 12 AB from Epoch 1 spanning the UV through optical. The binary components are marginally resolved in all filters. Each image is 40$\times$40 pixels, or about 1.56$\times$1.56\arcsec. North is up and East is left. Bottom: Co-added HST/WFC3-UVIS images of SR 12 c spanning the UV through optical. The signal-to-noise ratio of each of these detections are: 7.38, 12.7, 7.50, 10.5, and 108, for F225W, F336W, F438W, F555W, and F656N, respectively. Note the strong H$\alpha$ emission evident in F656N. The Balmer jump occurs at 3646 \AA\ and is noticeable between F336W and the weaker emission in F438W. Each image is 40$\times$40 pixels, or about 1.56$\times$1.56\arcsec. North is up and East is to the left.} 
    \label{fig:tile}
\end{figure*} 
\endgroup

\begingroup 
\nolinenumbers
\begin{figure}
    \centering
    \includegraphics[width =\columnwidth]{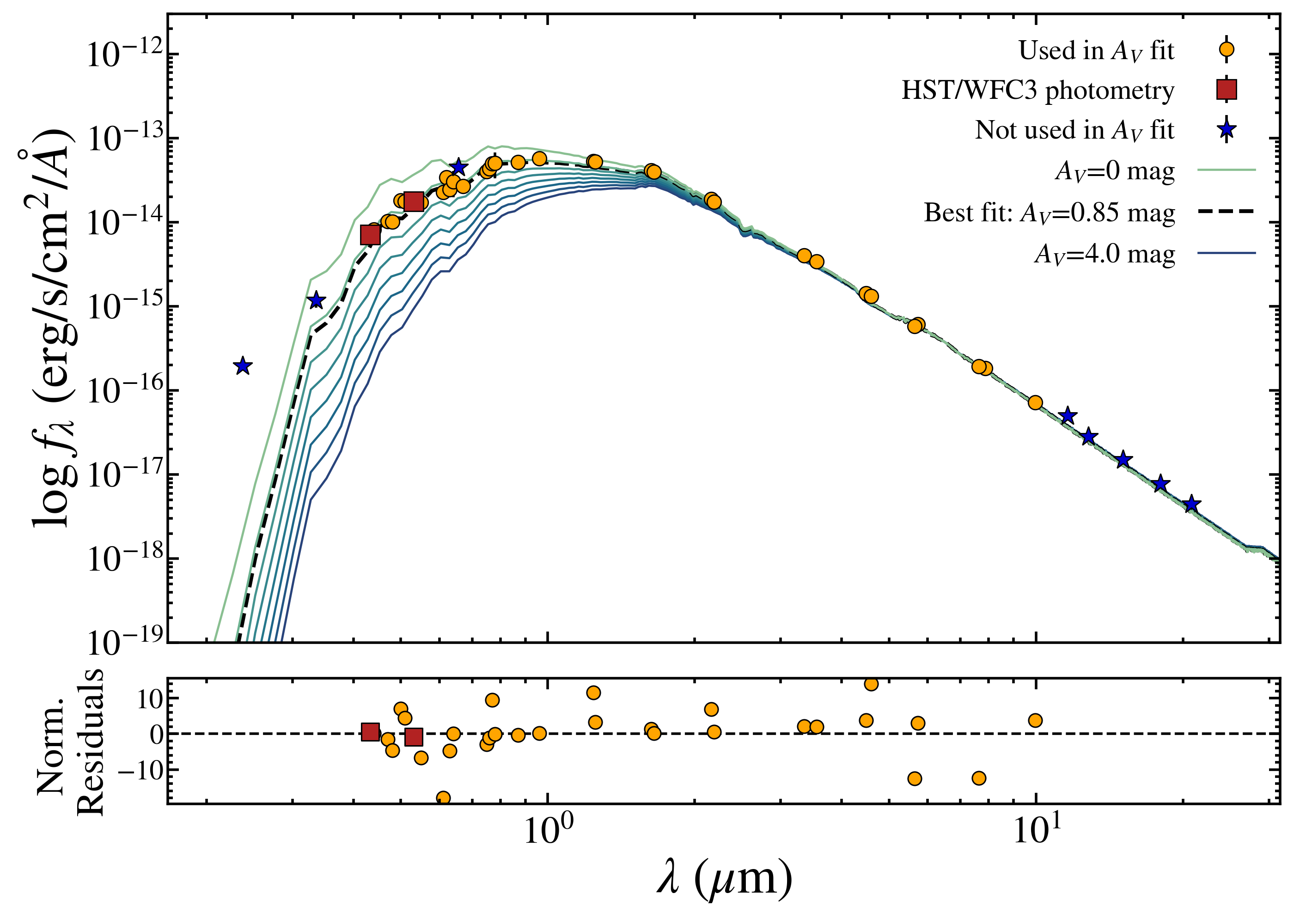}
    \caption{The combined SED for SR 12 AB described by a BT-Settl model at $T_\mathrm{eff}$ of 3800 K.  Interstellar reddening ranging from $A_V$=0 to 4 mag is shown with 0.5 mag spacing. The best fit yields an extinction of $A_V$=0.85, which is what we adopt for the host and its distant companion, SR 12 c.} 
    \label{fig:fanext}
\end{figure} 
\endgroup

\subsection{Extinction}\label{sec:ext}
\cite{Kuzuhara2011} compared a young M9 standard spectrum to the extinction-corrected spectrum of SR 12 c and reported a \textit{J-}band extinction $A_J$ of 0.5 $\pm$ 0.2 mag. \cite{Bowler2014} converted this to an $A_V$ of $1.7 \pm 0.7$ mag using the $A_J$/$A_V$ relationship from \cite{Rieke1985}. \cite{Martinez2022} and \cite{Wu2022} adopted this value, while \cite{SantamariaMiranda2018} made their own measurement of extinction from their X-shooter spectrum, letting $A_V$ vary as a free parameter in their photospheric model fits, yielding a value of 1.24 mag. Determining an accurate extinction is especially important for correctly establishing the accretion luminosity and mass accretion rate for SR 12 c, as the UV photometry will be most sensitive to the precision and accuracy of the inferred $A_V$. Given the various previous estimates of extinction, each with a large uncertainty, here we conducted our own investigation of the extinction to this system.

Our approach involves fitting a photosphere model from the BT-Settl \citep{Allard2012} grid to the SED of the host binary, SR 12 AB, and allowing extinction to vary. Given the comparable brightness and colors of SR 12 A and B \citep{Simon1987, Wilking2005}, we model them as a single unresolved source with one photospheric temperature, representing a star with properties between the two binary components. We compile published unresolved photometry of the binary components spanning 0.44 to 11.6 $\mu$m. We then measure unresolved HST/WFC3-UVIS aperture photometry for the binary following the same methodology we used for the companion (see Section~\ref{sec:hst} below). Combining the new photometry with published values, as well as JWST/MIRI photometry from Wu et al. (submitted), we assemble a UV through mid-IR SED (shown in  Figure~\ref{fig:fanext}) for the binary, with photometric values tabulated in Table~\ref{tab:binphot}. Adopting the \cite{LuhmanRieke1999} unresolved spectral type of M0 for SR 12 AB, we use the relations between spectral type and $T_\mathrm{eff}$ from \cite{Pecaut2013} to derive an estimated $T_\mathrm{eff}$ of 3850 K. For this work we adopt 3800 K for the photosphere model and a log $g$ of 4.0, as the grid spacing of the models is 100~K in $T_\mathrm{eff}$ and 0.5 dex in log $g$. The results are consistent within uncertainties if we instead select the model at 3900 K. 

Next, we generate synthetic photometry for each of 33 photometric points using the {\fontfamily{pcr}\selectfont species} package \citep{Stolker2020}. Values blueward of 3000 \AA\ and redward of 10 $\mu$m are removed to avoid the potential impact of chromospheric or accretion excess at short wavelengths and weak disk emission at long wavelengths. The H$\alpha$ photometry (F656N) is also removed, as it displays mild emission above the photosphere and would therefore bias the model fit. We then de-redden the synthetic photometry using the \cite{Fitzpatrick99} reddening law for $A_V$ values from $A_V = $ 0.0--4.0 in increments of 0.01 mag, and scale each extincted model data set to the observed data (Fig.~\ref{fig:fanext}). Finally, we compute the $\chi ^2$ value to assess the goodness of fit between the observed photometry and the extincted, synthetic model for SR 12 AB:

\begin{equation}\label{eq:chi}
    \chi ^2 = \sum_{i=1}^N\Big(\frac{d_i-m_i}{\sigma_i}\Big)^2
\end{equation}

\noindent where $d_i$ is the is the measured photometry for filter $i$, $m_i$ is the model synthetic photometry, and $\sigma_i$ represents the photometric uncertainties. The minimum $\chi ^2$ value is $\sim1500$ and corresponds to an extinction value of $A_V$ = 0.85~mag. The reduced $\chi ^2$ value, 45, is equivalent to $\chi ^2$/$\nu$, where $\nu$ = 31 is the number of data points (33) minus the number of model free parameters (2, a scale factor and the extinction). The reduced $\chi ^2$ value remains relatively large owing to the small uncertainties associated with the reported photometry.

Several studies have independently measured the (unresolved) spectral type of SR 12 AB to be M0 (e.g., \citealt{LuhmanRieke1999, Wahhaj2010, Pecaut2016}). We therefore adopt an uncertainty of 0.5 subtypes for the (unresolved) spectrum of SR 12 AB.  This corresponds to upper and lower uncertainties of +15 K and --85 K, respectively \citep{Pecaut2013}. The BT-Settl models are sampled with a grid spacing of 100~K, so the 3700~K and 3900~K models (about twice the implied effective temperature range) are initially chosen for this exercise. This results in an extinction of 0.85 $\pm$ 0.34 mag; however, we halve these uncertainties to reflect the higher level of precision in spectral classification and adopt a final $A_V$ of 0.85 $\pm$ 0.17 mag for both SR 12 AB and SR 12 c. This is slightly lower than previous estimates, but higher extinction values are incompatible with the observed SED and under-predict the photometry of SR 12 AB at short wavelengths. Though lower, this value is still within $2\sigma$ of previous estimates.

To assess how the assumption of a single-component model impacts our estimate of $A_V$, we use the resolved $K$-band photometry from \cite{Simon1987} along with evolutionary models from \cite{Baraffe2015} to construct a two-component model with effective temperatures of 3800~K for SR 12 A and 3700~K for B. After adjusting the relative scaling of each component based on the $K$-band flux ratio and summing the two photospheres, we find a best-fit extinction of 0.87~mag---nearly identical to our single-component model.

Extinction values in the $\rho$ Oph cloud and the L1688 cloud core in particular can range from $A_V$=0 to as high as $A_V$=20 mag \citep{Canovas2019}. \cite{Rizzuto2015} found that for a population of 20 members of the $\rho$ Oph complex without detected disks, the average extinction is $A_V = 0.9\pm0.6$ mag. Our estimate of 0.85 $\pm$ 0.17 mag therefore agrees well with other non-embedded members of $\rho$ Oph that, like SR 12 AB, do not have disks. Because extinction does not significantly vary in this region over a few arcseconds \citep{Canovas2019}, we assume SR 12 c shares the same $A_V$ as SR 12 AB. Additionally, we do not consider local reddening at the source and assume the CPD of SR 12 c is not contributing to extinction.

\begingroup 
\nolinenumbers
\begin{deluxetable*}{ccccc}
\renewcommand\arraystretch{0.9}
\tabletypesize{\small}
\setlength{\tabcolsep}{0.1cm}
\tablewidth{0pt}
\tablecolumns{4}
\tablecaption{Unresolved Photometry of SR 12 AB \label{tab:binphot}}
\tablehead{
\colhead{Filter} & \colhead{Central Wavelength} & \colhead{Observed Flux Density} & \colhead{De-reddened Flux Density$^a$} & Reference \\
\colhead{} & \colhead{($\mu$m)} & \colhead{($10^{-15}$ erg s$^{-1}$ cm$^{-2}$ \AA$^{-1}$)} & \colhead{($10^{-15}$ erg s$^{-1}$ cm$^{-2}$ \AA$^{-1}$)}}
\startdata
WFC3/F225W & 0.237 & 0.20 $\pm$ 0.01 & 1.47 $\pm$ 0.7 & This work \\
WFC3/F336W & 0.336 & 1.2 $\pm$ 0.2 & 4.39 $\pm$ 1.5 & This work \\
WFC3/F438W & 0.433 & 7.3 $\pm$ 0.7 & 20.5 $\pm$ 4.9 & This work \\
Johnson/$B$ & 0.440 & 8.1 $\pm$ 0.7 & 22.6 $\pm$ 2.0 & \cite{Henden2016} \\
POSS-II/$J$ & 0.470 & 10.2 $\pm$ 0.5 & 26.5 $\pm$ 1.4 & \cite{Lasker2008} \\
PAN-STARRS1/$g$ & 0.477 & 13.1 $\pm$ 0.3 & 25.6 $\pm$ 0.8 & \cite{Chambers2016} \\
Gaia DR3/$G_{BP}$ & 0.500 & 18.1 $\pm$ 0.2 & 43.5 $\pm$ 0.6 & \cite{Gaia2016b, Gaia2023j} \\
Gaia DR2/$G_{BP}$ & 0.510 & 17.6 $\pm$ 0.2 & 41.3 $\pm$ 0.6 & \cite{Gaia2016b, Gaia2018b} \\
WFC3/F555W & 0.531 & 18 $\pm$ 2 & 39.4 $\pm$ 7.62 & This work \\
PAN-STARRS1/$r$ & 0.610 & 22.8 $\pm$ 0.3 & 44.3 $\pm$ 0.6 & \cite{Chambers2016} \\
Gaia DR2/$G$ & 0.620 & 34.1 $\pm$ 0.1 & 65.2 $\pm$ 0.2 & \cite{Gaia2016b, Gaia2018b} \\
SDSS/$r$ & 0.630 & 24.5 $\pm$ 0.9 & 46.1 $\pm$ 1.7 & \cite{Howell2014} \\
POSS-II/$F$ & 0.640 & 30.3 $\pm$ 1.1 & 56.3 $\pm$ 19.5 & \cite{Lasker2008} \\
WFC3/F656N & 0.656 & 45 $\pm$ 5 & 81.1 $\pm$ 13 & This work \\
PAN-STARRS1/$i$ & 0.750 & 40 $\pm$ 3 & 65.1 $\pm$ 4.3 & \cite{Chambers2016} \\
SDSS/$i'$ & 0.760 & 43 $\pm$ 5 & 69.6 $\pm$ 8.3 & \cite{Henden2016} \\
Gaia DR2/$G_{RP}$ & 0.770 & 49.6 $\pm$ 0.3 & 78.7 $\pm$ 0.5 & \cite{Gaia2016b, Gaia2018b} \\ 
POSS-II/$i$ & 0.780 & 50.2 $\pm$ 0.2 & 79.0 $\pm$ 26.8 & \cite{Lasker2008} \\
PAN-STARRS1/$z$ & 0.870 & 52 $\pm$ 9 & 75.9 $\pm$ 12.7 & \cite{Chambers2016} \\
PAN-STARRS1/$y$ & 0.960 & 57 $\pm$ 9 & 78.5 $\pm$ 12.0 & \cite{Chambers2016} \\
2MASS/$J$ & 1.24 & 53.4 $\pm$ 0.4 & 65.5 $\pm$ 0.5 & \cite{Marsh2010} \\
Johnson/$J$ & 1.25 & 52.6 $\pm$ 1.2 & 64.3 $\pm$ 1.4 & \cite{Cutri2003} \\
Johnson/$H$ & 1.63 & 41.4 $\pm$ 1.7 & 47.2 $\pm$ 1.9 & \cite{Cutri2003} \\
2MASS/$H$ & 1.65 & 39.7 $\pm$ 1.7 & 45.2 $\pm$ 1.9 & \cite{Evans2003} \\
2MASS/$K_s$ & 2.16 & 18.9 $\pm$ 0.1 & 20.6 $\pm$ 0.1 & \cite{Marsh2010} \\
Johnson/$K$ & 2.19 & 17.5 $\pm$ 0.3 & 19.1 $\pm$ 0.3 & \cite{Cutri2003} \\
WISE/W1 & 3.35 & 3.98 $\pm$ 0.08 & 4.20 $\pm$ 0.08 & \cite{Marocco2021} \\
IRAC/3.6 $\mu$m & 3.55 & 3.42 $\pm$ 0.07 & 3.60 $\pm$ 0.07 & \cite{Gunther2014} \\
IRAC/4.5 $\mu$m & 4.49 & 1.43 $\pm$ 0.03 & 1.50 $\pm$ 0.03 & \cite{Gunther2014} \\
WISE/W2 & 4.60 & 1.33 $\pm$ 0.01 & 1.40 $\pm$ 0.01 & \cite{Marocco2021} \\
MIRI/F560W & 5.64 & 5.840 $\pm$ 0.003 & 0.600 $\pm$ 0.003 & Wu et al. (submitted) \\
IRAC/5.8 $\mu$m & 5.73 & 0.620 $\pm$ 0.005 & 0.600 $\pm$ 0.010 & \cite{Gunther2014} \\
MIRI/F770W & 7.64 & 0.190 $\pm$ 0.009 & 0.200 $\pm$ 0.001 & Wu et al. (submitted) \\
IRAC/8.0 $\mu$m & 7.87 & 0.190 $\pm$ 0.002 & 0.200 $\pm$ 0.002 & \cite{Gunther2014} \\
MIRI/F1000W & 9.95 & (70 $\pm$ 0.8)$\times 10^{-4}$ & (100 $\pm$ 2)$\times 10^{-4}$ & Wu et al. (submitted) \\
WISE/W3 & 11.6 & (50 $\pm$ 0.2)$\times 10^{-4}$ & (100 $\pm$ 10)$\times 10^{-4}$ & \cite{Stassun2019} \\
MIRI/F1280W & 12.8 & (28.1 $\pm$ 1.4)$\times 10^{-4}$ & (28.5 $\pm$ 1.4)$\times 10^{-4}$ & Wu et al. (submitted)\\
MIRI/F1500W & 15.1 & (15.1 $\pm$ 1.0)$\times 10^{-4}$ & (15.3 $\pm$ 1.0)$\times 10^{-4}$ & Wu et al. (submitted)\\
MIRI/F1800W & 18.0 & (7.80 $\pm$ 0.70)$\times 10^{-4}$ & (7.87 $\pm$ 0.70)$\times 10^{-4}$ & Wu et al. (submitted)\\
MIRI/F2100W & 20.8 & (4.45 $\pm$ 0.40)$\times 10^{-4}$ & (4.48 $\pm$ 0.40)$\times 10^{-4}$ & Wu et al. (submitted)\\
\enddata
\tablecomments{$^a$Observed photometry de-reddened by $A_V=0.85$ $\pm$ 0.17 mag.}
\end{deluxetable*} 
\endgroup

\subsubsection{HST Photometry}\label{sec:hst}
We began by downloading the {\fontfamily{pcr}\selectfont drc} HST/WFC3-UVIS FITS files from the Mikulski Archive for Space Telescopes (MAST) at the Space Telescope Science Institute\footnote{All observations of the SR 12 system analyzed in this work can be accessed via\dataset[doi:10.17909/fz8z-yk57]{https://doi.org/10.17909/fz8z-yk57}.}. The {\fontfamily{pcr}\selectfont drc} extension denotes that the frames have been processed through AstroDrizzle \citep{Fruchter2010}, the Python-based pipeline automating the cosmic-ray removal and subsequent aligning and combining of WFC3 images from each dither, as well as correcting for geometric distortion and charge-transfer efficiency. In this case, 3 dithers were combined to create one {\fontfamily{pcr}\selectfont drc} frame. The host binary is marginally resolved in all frames with an elongation largely in the east-west direction. SR 12 c is visually evident in all frames, but is rather faint in the UV filters (F225W and F336W). To compensate for this, we co-add the frames for the two epochs ($\sim$1 month apart) in each of the 5 filters. Later, in Section~\ref{sec:var}, we explore photometric variability across this month-long time baseline between each epoch. 

Our approach to precisely register both north-aligned images is to cross correlate the processed frames from February (``A" frames) with those from March (``B"). The cross-correlation between the A and B frames is computed using the {\fontfamily{pcr}\selectfont phase\_cross\_correlation} routine in the Python {\fontfamily{pcr}\selectfont scikit-image} package to determine the $x$ and $y$ positional offsets between the two frames. This is driven by the position of SR 12 AB, which is the brightest object in each image. No image rotation is applied, as each frame has the same position angle of reference aperture center as determined by the header keyword {\fontfamily{pcr}\selectfont ORIENTAT}, reported in degrees East of North. We perform a sub-pixel shift using the {\fontfamily{pcr}\selectfont scipy} function {\fontfamily{pcr}\selectfont ndimage.fourier\_shift} to align the images and co-added the frames into one combined image for each filter. 

The FITS files are converted to flux density in units of erg s$^{-1}$ cm$^{-2}$ \AA$^{-1}$ via the inverse sensitivity conversion factors in the {\fontfamily{pcr}\selectfont PHOTFLAM} FITS header keyword. Aperture photometry is then carried out with the {\fontfamily{pcr}\selectfont photutils} package on each co-added frame using an aperture radius of 2 times the PSF FWHM at a given wavelength, corresponding to 3.94 pix for F225W, 3.61 pix for F336W, 3.41 pix for F438W, 3.36 pix for F555W, and 3.43 pix for F656N. The FWHM decreases from the UV to the blue-optical filters because the near-UV filters are impacted by imager polishing errors, which transfer power from the PSF core to the non-Gaussian PSF wings, widening and strengthening the wings and the FWHM.

To generate realistic, empirically based uncertainties, we carry out injection-recovery tests of synthetic point spread functions (PSFs) at various locations in the science image. We adopted HST/WFC3 images of FU Tau A (also from GO 16302\footnote{The specific observations of FU Tau used for the PSF templates can be accessed via\dataset[doi:10.17909/m9fx-nk12]{https://doi.org/10.17909/m9fx-nk12}.}) as PSF templates by scaling them to the amplitude of the SR 12 c PSF in each image, and then injecting 100 versions of this mock PSF---each with randomly-generated Poisson noise---into the science frame of SR 12 c. In this way, our synthetic PSFs included the dominant sources of uncertainty that are impacting the photometry of SR 12 c: Poisson noise, read noise, and sky background. Then, we perform aperture photometry on each of these 100 instances of the mock PSF using the same aperture radii used for the companion and adopted the sample standard deviation as an estimate of the photometric uncertainty for SR 12 c. These empirical uncertainties are about an order of magnitude higher than the photon noise.

To account for the flux missed by the finite aperture applied to SR 12 c (3--4 pixels), we first corrected to the ``standard" WFC3 aperture radius of 10 pixels, beyond which the encircled energy fraction does not significantly vary with time or position on the detector \citep{Calamida2021}. This adjustment is carried out using empirically calculated correction factors which we measure across multiple filters using other targets from this broader HST program (GO 16302): CT Cha A, FU Tau A, SCH06-J0359+2009 A, and the isolated planetary-mass object OTS-44\footnote{The observations of the four objects used for aperture corrections can be accessed via \dataset[doi:10.17909/m9fx-nk12]{https://doi.org/10.17909/m9fx-nk12}.}. As these targets were observed across multiple years, they sample a variety of PSF realizations under differing thermal conditions of the observatory and with positions at several locations on the detector. We assume an aperture radius the same size as applied to SR 12 c, as well as a 10-pixel aperture, and compute the ratio between the flux in the smaller aperture and the larger aperture. The mean and standard deviation of these ratios among the four calibration targets is used as a multiplicative correction factor, ranging from 1.04 to 1.21 for the five filters. We then multiply the raw photometry for SR 12 c in each filter by the appropriate correction factor and propagate the flux uncertainty. To extend the aperture from 10 pixels out to infinity, and thus capture the total flux of SR 12 c, we implement the corrections tabulated in the WFC3 instrument documentation \citep{Marinelli2024}. The resulting photometry is listed in Table~\ref{tab:hstphotlog} and shown in Figure~\ref{fig:sed}. 

To de-redden the photometry, we employ the dust extinction function from \cite{Fitzpatrick99} via the Python {\fontfamily{pcr}\selectfont extinction} package \citep{Barbary2016} using an $R_V$ of 3.1. Uncertainties are propagated in a Monte Carlo fashion for 10$^6$ trials assuming an extinction of $A_V$ = 0.85 $\pm$ 0.17 mag (see Section~\ref{sec:ext}). We adopt the mean and standard deviation of this distribution as the final flux and corresponding error. These values are displayed in Table~\ref{tab:hstphotlog} and shown in Figure~\ref{fig:sed}.

This methodology is repeated to determine photometry for SR 12 AB, which is treated as an unresolved single source, beginning from the standardized 10 pixel aperture radius. Considering the strong signal from the binary, co-adding frames across the two epochs is not necessary. Instead, we report the weighted average between the two epochs (see Table~\ref{tab:hstphotlog}).

\begingroup 
\nolinenumbers
\begin{figure*}[t]
    \centering
    \includegraphics[width =\textwidth]{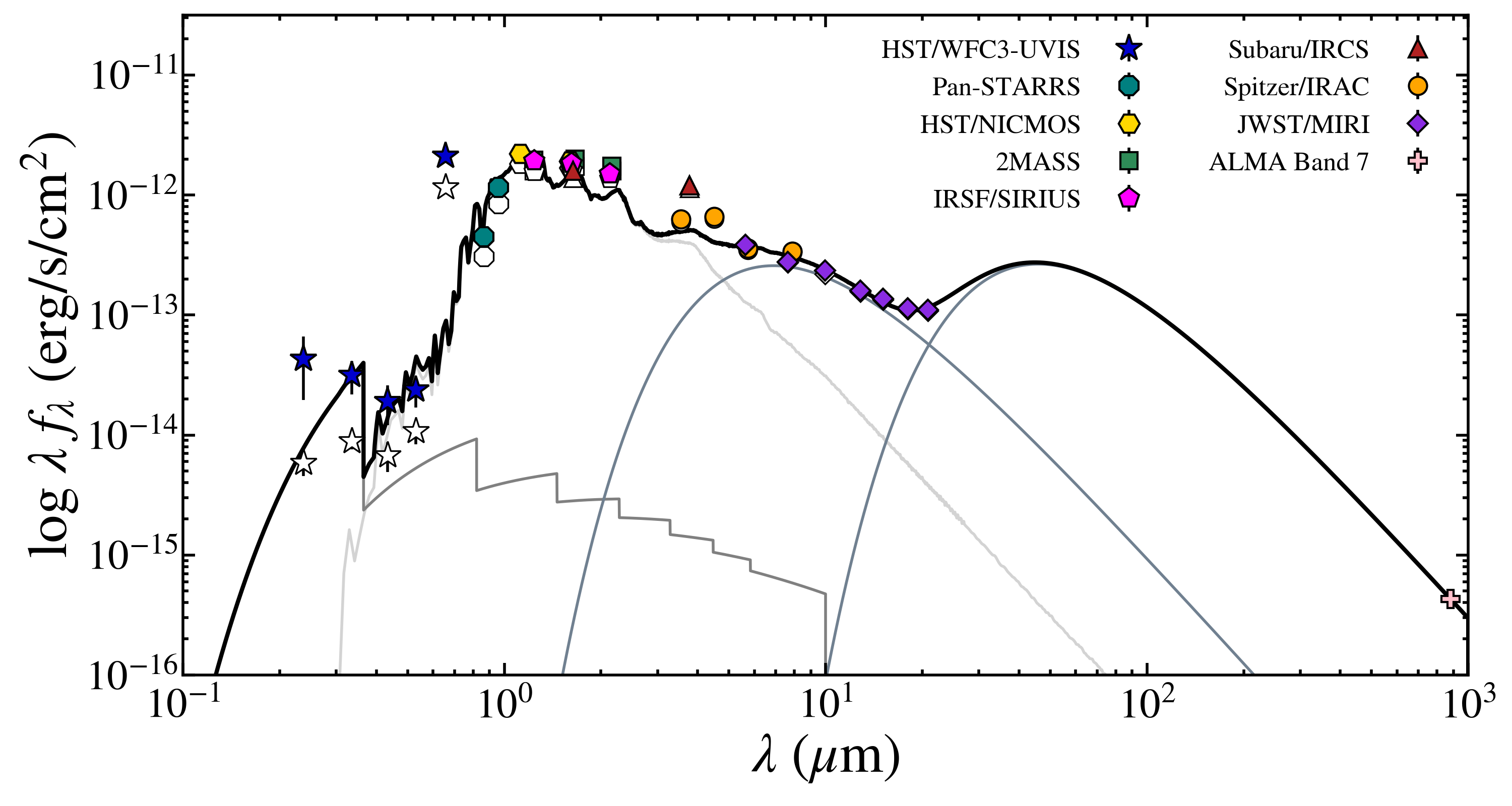}
    \caption{The 0.2--880 $\mu$m SED of SR 12 c, with three primary sources of emission: accretion in the UV, the photosphere in the near-infrared, and a circumplanetary disk spanning the mid-IR to mm wavelengths. In the UV, the best-fitting hydrogen slab model \citep{Valenti1993, Herczeg2008} is shown, with a temperature of 10640 K and a density of 2.97$\times 10^{13}$ cm$^{-3}$ (dark gray). In the near-IR, an atmospheric model is shown \citep{Allard2012} with an effective temperature of 2600 K and a log $g$ of 3.5 (light gray). In the mid-IR through sub-mm, a two-blackbody disk model describes the emission, with temperatures of 533 and 80 K, respectively (blue-gray). The black line shows the summation of all of the models. The complete photometry is described in Table~\ref{tab:sedphot}, where the de-reddened photometry is shown here as a colored symbol and the observed photometry is shown as an open, white symbol.} 
    \label{fig:sed}
\end{figure*}
\endgroup

\begingroup 
\nolinenumbers
\begin{deluxetable}{ccccccc}
\renewcommand\arraystretch{0.9}
\tabletypesize{\small}
\setlength{\tabcolsep}{0.1cm}
\tablewidth{0pt}
\tablecolumns{11}
\tablecaption{HST/WFC3-UVIS Observing Log \label{tab:hstobs}}
\tablehead{
\colhead{UT Date} & \colhead{Filter} & \colhead{$\lambda_{\mathrm{eff}}$} & \colhead{Exp. Time} & \colhead{SNR} \\
\colhead{(YYYY-MM-DD)} & \colhead{} & \colhead{(\AA)} & \colhead{(s)} & \colhead{}
} 
\startdata
2021-02-28 & F225W & 2372.1 & 774 & 5.87\\
$\cdots$ & F336W & 3354.5 & 291 & 13.0\\ 
$\cdots$ & F438W & 4326.2 & 156 & 4.63\\
$\cdots$ & F555W & 5308.4 & 45 & 6.40\\
$\cdots$ & F656N & 6561.4 & 516 & 86.0\\
2021-03-26 & F225W & 2372.1 & 774 & 6.07\\
$\cdots$ & F336W & 3354.5 & 291 & 6.34\\
$\cdots$ & F438W & 4326.2 & 156 & 6.42\\
$\cdots$ & F555W & 5308.4 & 45 & 8.87\\
$\cdots$ & F656N & 6561.4 & 516 & 66.6\\
\enddata
\tablecomments{All observations are carried out using a 3-point ``line" dither pattern and a point spacing of 0$\farcs$135 with 3 individual exposures per filter.}
\end{deluxetable}
\endgroup

\begingroup 
\nolinenumbers
\begin{deluxetable*}{lccccc} 
\renewcommand\arraystretch{0.9}
\tabletypesize{\small}
\setlength{ \tabcolsep } {.1cm}
\tablewidth{0pt}
\tablecolumns{4}
\tablecaption{HST/WFC3 Photometry of the SR 12 system \label{tab:hstphotlog}}
\tablehead{
 \colhead{Filter} & \colhead{SR 12 AB $f_\lambda$} & \colhead{SR 12 AB $f_\lambda$ (Ext. Cor.)$^{a}$} & \colhead{SR 12 c $f_\lambda$} & \colhead{SR 12 c $f_\lambda$ (Ext. Cor.)$^{a}$}\\
 \colhead{} & \colhead{(10$^{-16}$ erg s$^{-1}$ cm$^{-2}$ \AA$^{-1}$)} & \colhead{(10$^{-16}$ erg s$^{-1}$ cm$^{-2}$ \AA$^{-1}$)} & \colhead{(10$^{-18}$erg s$^{-1}$ cm$^{-2}$ \AA$^{-1}$)} & \colhead{(10$^{-18}$ erg s$^{-1}$ cm$^{-2}$ \AA$^{-1}$)} 
}
\startdata
\multicolumn{5}{c}{Epoch 1 (2021-02-28)} \\
\tableline
F225W$^{b, }{^c}$ & $2.09 \pm 0.032$  & $15.7 \pm 7.18$ & $2.34 \pm 0.51$ & $17.2 \pm 9.22$\\
F336W             & $14.4 \pm 0.017$ & $51.5 \pm 13.8$  & $3.77 \pm 0.37$  & $13.5 \pm 3.89$\\
F438W             & $79.3 \pm 0.072$ & $225 \pm 48.4$   & $1.55 \pm 0.32$ & $4.36 \pm 1.35$\\
F555W             & $192 \pm 0.53$  & $430 \pm 71.0$    & $1.73 \pm 0.37$  & $3.85 \pm 1.06$\\
F656N             & $493 \pm 1.34$  & $894 \pm 107$     & $201 \pm 3.1$      & $365 \pm 44.4$\\
\hline
\multicolumn{5}{c}{Epoch 2 (2021-03-26)} \\
\tableline
F225W             & $1.84 \pm 0.035$  & $13.8 \pm 6.32$   & $2.54 \pm 0.54$ & $18.7 \pm 9.2$\\
F336W             & $9.90 \pm 0.014$  & $35.5 \pm 9.53$   & $1.89 \pm 0.39$ & $6.69 \pm 2.35$\\
F438W             & $64.8 \pm 0.059$  & $184 \pm 39.5$    & $1.82 \pm 0.39$ & $5.10 \pm 1.60$\\
F555W             & $157 \pm 0.43$  & $351 \pm 57.9$      & $2.31 \pm 0.39$ & $5.12 \pm 1.24$\\
F656N             & $392 \pm 1.06$  & $712 \pm 86.1$      & $153 \pm 3.2$   & $278 \pm 34.0$\\
\hline
\multicolumn{5}{c}{Co-added Photometry$^{d}$} \\ 
\hline
F225W            & $1.96 \pm 0.13$  & $14.7 \pm 6.81$   & $2.47 \pm 0.56$  & $18.1 \pm 9.75$\\
F336W            & $12.4 \pm 2.21$  & $44.0 \pm 14.6$   & $2.64 \pm 0.37$  & $9.39 \pm 2.92$\\
F438W            & $72.8 \pm 7.22$  & $206  \pm 49.4$   & $1.57 \pm 0.43$  & $4.39 \pm 1.59$\\
F555W            & $177  \pm 17.6$  & $393 \pm 76.5$    & $2.02 \pm 0.45$  & $4.49 \pm 1.26$\\
F656N            & $448  \pm 49.9$  & $811 \pm 134$     & $177  \pm 3.3$   & $322  \pm 39.3$\\
\enddata
\tablecomments{$^a$Observed photometry de-reddened by $A_V$= 0.85 $\pm$ 0.17 mag.} 

\tablecomments{$^b$The observed F225W photometry exceeds the expected contribution from the photosphere by over 4 orders of magnitude and is dominated by accretion excess. As such, we do not apply any correction for potential contamination from the WFC3 red leak \citep{Brown2008}.}

\tablecomments{$^c$The F225W photometric point has particularly large error bars due in part to the impact of the 2175 \AA\ extinction bump, which overlaps with the wavelength range of the F225W filter.}

\tablecomments{$^d$For SR 12 AB, this measurement is a weighted average between the two epochs. For SR 12 c, this measurement comes from the co-added image.} 
\end{deluxetable*}
\endgroup

\subsection{Chance Alignment Analysis}\label{sec:cluster}
Given the large projected separation between SR 12 AB and c (8$\farcs$7), it is possible that they represent a chance alignment of two nearby cluster members and are not gravitationally bound. From the density of stars in the L1688 cloud core (125 sources within within 0$\fdg$6 from the extinction peak, \citealt{Canovas2019}), we measured a local surface density of field sources of $\Sigma = 110.52 ~\mathrm{deg^{-2}}$. The probability of coincidental alignment in the cloud is computed within three circular regions, with $r=1.2\times10^{3}$ AU (the separation of SR 12 c), $r=5\times10^{3}$ AU, and $r=10^{4}$ AU, the ultrawide binary tail where the separation distribution of wide field binaries steepens and chance alignments are expected to become much more common \citep{ElBadry2021}. Using Poisson statistics, the probability of finding at least one unrelated source in the aperture is $P(\ge1)=1-e^{-\lambda}$, where $\lambda$ is the rate of chance alignment. The chance-alignment probability is 0.2\% at $r=1.2\times10^{3}$ AU, 3.4\% at $r=5\times10^{3}$ AU, and 13.0\% at $r=10^{4}$ AU. Considering the low probability at all separations, notably below 1\% at the separation of SR 12 c, we proceeded with the assumption that SR 12 c is indeed bound to its host binary.

The nearest member to the SR 12 system is YLW 13B (2MASS 16272146–2441430), an embedded young stellar object in the L1688 cloud with a dense DCO$^+$ core \citep{Loren1990}. When repeating the chance alignment statistical test for an object at the separation of YLW 13B ($\sim$3500 AU), we found a 1.65\% chance of random alignment. While this indicates that it is fairly unlikely for a cluster member like YLW 13B to have fallen so close to SR 12 AB by chance, it is still an order of magnitude more likely than the probability of SR 12 c being aligned with the host binary by chance. It is possible that there is a physical relationship between SR 12 ABc and YLW 13B given that the binary semimajor axis distribution extends to 10$^4$ AU \citep{Raghavan2010}, or perhaps that the space density is somewhat higher in this portion of the $\rho$ Oph complex.

\section{Modeling the SED of SR 12 \lowercase{c}} \label{sec:mods}
To probe the accretion properties of SR 12 c, we independently fit models of hot hydrogen gas to the UV photometry and atmospheric models to the optical and infrared region. At longer wavelengths, beyond 5 $\mu$m, disk emission dominates and is well-described with a two-component blackbody. Figure~\ref{fig:sed} shows the full SED for SR 12 c spanning 0.2 $\mu$m to 880 $\mu$m. It includes observations from HST (this work), ground-based facilities \citep{Allen2002, Cutri2003, Kuzuhara2011}, \textit{Spitzer} \citep{Martinez2022}, JWST (Wu et al., submitted), and ALMA \citep{Wu2022}. To our knowledge, this represents the broadest wavelength coverage of any planetary-mass object to date. Table~\ref{tab:sedphot} lists all the photometry used to construct the SED. Here, we describe details of each component separately and how the models are fit to the observed SED. 

\subsection{Photosphere Models}\label{sec:photos}
Thermal emission from SR 12 c dominates its SED at near-infrared wavelengths. For this region, we use atmospheric models from the BT-Settl grid \citep{Allard2012} and normalize its peak to the location of the reported SIRIUS \textit{J-}band photometry, which represents a balance between accretion excess at bluer wavelengths and disk emission at longer wavelengths. \cite{SantamariaMiranda2018} reported that the best model atmosphere fits to their X-shooter spectrum of SR 12 c fall in the range of 2500–2800 K with log $g$ between 3.5 and 4.5. For this work we adopt a solar metallicity 2600 K model with a surface gravity of 3.5, as this is most consistent with model predictions given the bolometric luminosity of SR 12 c and an age of 3-4 Myr (e.g. \citealt{Baraffe2015, Phillips2020}). Note that the photospheric contribution is negligible shortward of $\sim$4000 \AA\ and longward of $\sim$7 $\mu$m and thus does not significantly change our interpretation of either the UV or mid-IR excess.

\subsection{Accretion Models}\label{sec:slab} 
The shortest-wavelength portion of the SED of SR 12 c is the UV through red-optical regime. The Balmer jump at 3646 \AA\ is a convenient diagnostic of mass accretion rate. It is created by a sharp rise in opacity at $\lambda$ $<$ 3645 \AA\ from photoionization of excited ($n$=2) levels in the hot, dense accretion shock region. Figure~\ref{fig:contour} shows the strength of the Balmer jump (quantified as the ratio of flux before and after the jump, at 3600 \AA\ and 3700 \AA, respectively) as a function of local hydrogen number density and gas temperature. Low densities and temperatures result in the largest Balmer jumps with flux ratios approaching 100 across the discontinuity, while high temperatures and densities produce the weakest breaks. 

To quantify the Balmer jump and the excess blue continuum emission, we utilize a hydrogen slab model as introduced by \cite{Valenti1993}. The simple hydrogen slab models are agnostic to the mechanism and geometry of accretion. However, assuming that the emission originates from hot, shock-excited gas in an accretion shock near the surface of the planet allows for a precise determination of the accretion luminosity, and hence the mass accretion rate. If the accretion shock region is optically thick, this approach is expected to be substantially more accurate than estimates through line emission alone (such as H$\alpha$) \citep{Herczeg2008, Ardila2013, Ingleby2013, Alcala2014, Alcala2017, Robinson2019}. However, UV wavelengths suffer more heavily from uncertainties in extinction. 

In these models, \cite{Valenti1993} assumed the post-shock emission could be approximated by an isothermal, plane-parallel slab of pure hydrogen in local thermal equilibrium. Free parameters include temperature ($T$), column density ($n_e$), path length of the emitting gas (effectively the gas opacity), turbulent velocity, and filling factor of the accretion slab. Increasing the volume of the slab modulates the relative proportions of the emission from the accreting area and emission from the host planet; and adjusting the length of the slab relative to its surface area affects the total optical depth through the slab. $T$ increases the kinetic energy of the hydrogen atoms in the UV continuum, directly impacting the size of the Balmer jump, while $n_e$ primarily affects the wings of the Balmer jump, changing the shape of the feature rather than its size. This breaks any potential degeneracy where the same SED could be described by \textit{either} higher $T$ or higher $n_e$. For this study, we focus on constraining the region surrounding the Balmer jump, which can then be used to determine the accretion flux and luminosity. 

\cite{Herczeg2009} found that two parameters, temperature and number density, are expected to dominate the shape and behavior of the slab models, along with a linear, multiplicative normalization constant adjusting the overall amplitude of the model. The path length of the slab does not significantly impact the accretion luminosity measurement, and the filling factor is absorbed by the linear scaling constant when used to adjust the amplitude of the slab model. We follow this same approach by only considering temperature, density, and the scaling constant. Figure~\ref{fig:tempdens} shows the effect of these parameters on the shape and behavior of the hydrogen slab emission spectrum.\footnote{While less physically realistic than accretion shock models (e.g., \citealt{Calvet1998}), the slab models do not rely on any particular accretion flow morphology (the details of which are unknown in this case) and are robust to account for different geometries or multicolumn flows for the emitting gas \citep{Pittman2025}. In addition, the slab models and shock models both provide similar bolometric corrections to attain the final accretion luminosity \citep{Zhou2014}.}

\begingroup 
\nolinenumbers
\begin{figure}
    \centering
    \includegraphics[width =\columnwidth]{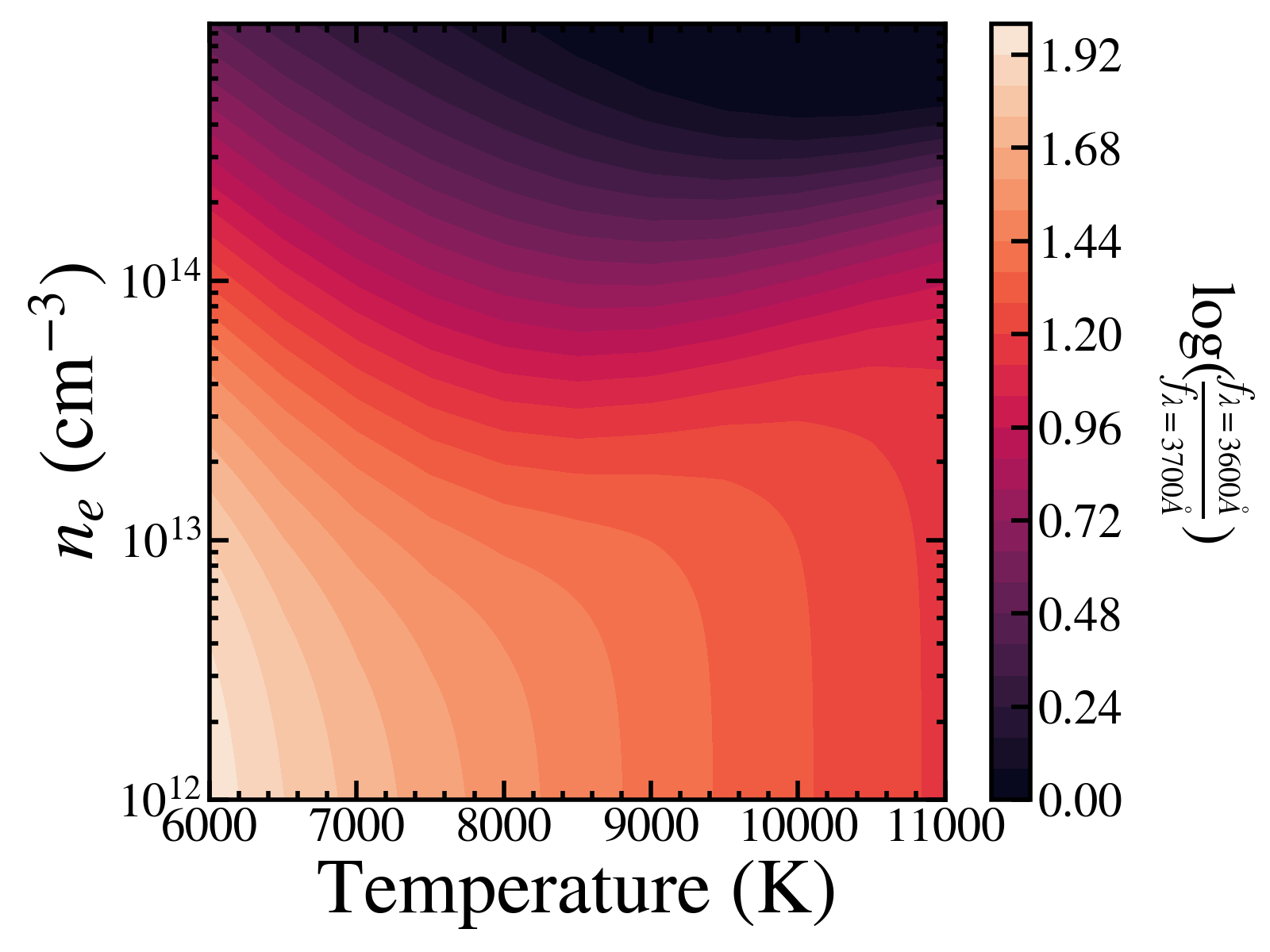}
    \caption{The strength of the Balmer jump modeled by the hydrogen slab as a function of column density ($n_e$) and temperature. The size of the jump is represented by the ratio of the flux coming from 3600 \AA\ to 3700 \AA. Note that the overall strength of the jump increases with temperature and number density.} 
    \label{fig:contour}
\end{figure} 
\endgroup

\begingroup 
\nolinenumbers
\begin{figure*}
    \centering
    \includegraphics[width =0.9\textwidth]{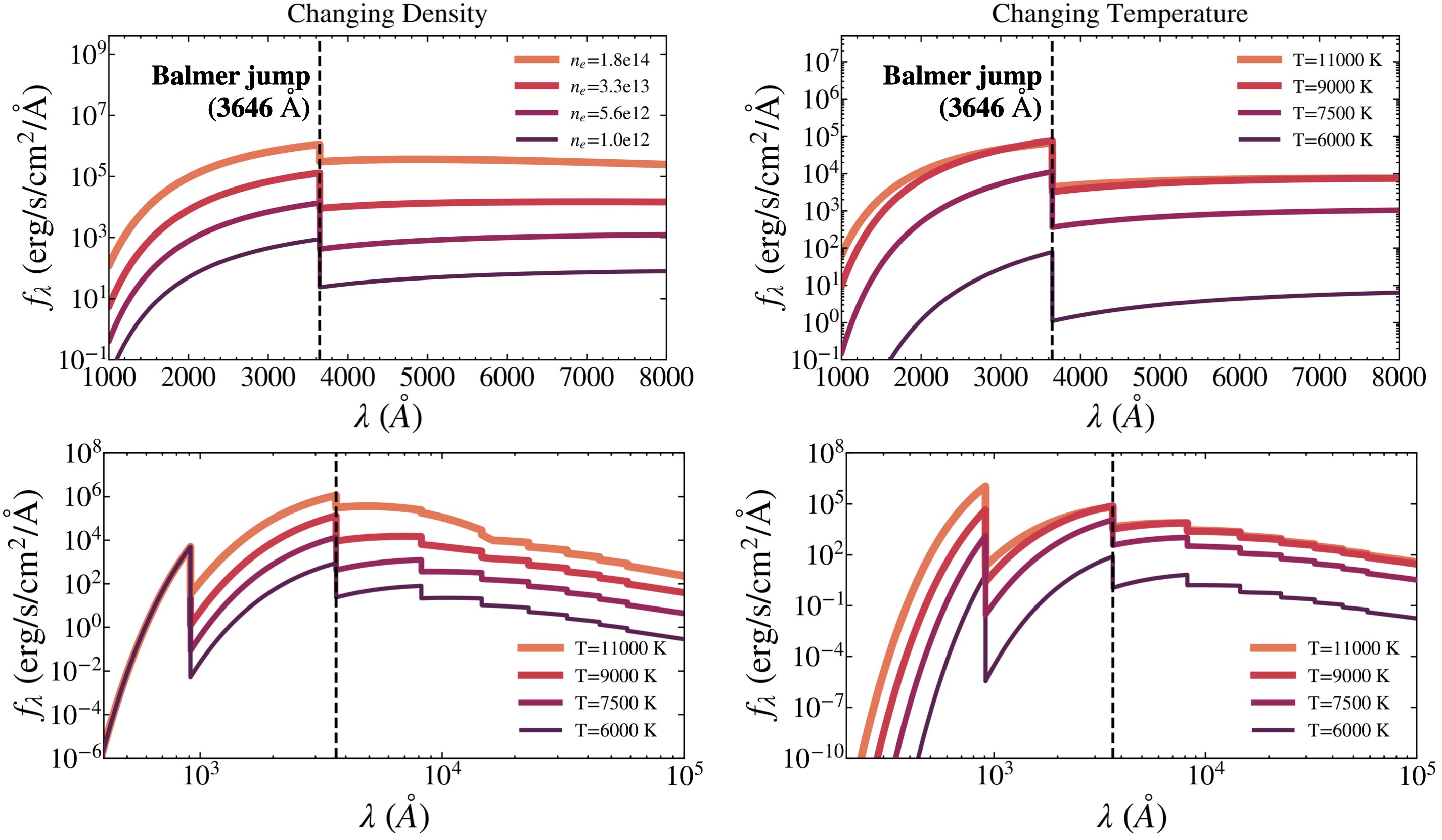}
    \caption{The effect of changing density and temperature on the size and shape of the Balmer jump in the hydrogen slab models. Upper left: At a fixed temperature ($T=8000$ K, in this case), varying number density affects the shape of the Balmer jump. Higher densities cause the slope on the red side of the break to flatten, the size of the Balmer jump to shrink, and the bolometric flux to increase. Bottom left: the same as upper left, but zoomed out to show the broader shape of the model. Upper right: At a fixed number density ($n_e=9.9\times10^{12}$ cm$^{-3}$, in this case), varying temperature primarily affects the strength of the Balmer jump. Higher temperatures result in a smaller jump and higher bolometric flux. Bottom right: the same as upper right, but zoomed out.} 
    \label{fig:tempdens}
\end{figure*} 
\endgroup

\begingroup 
\nolinenumbers
\begin{figure}
    \centering
    \includegraphics[width=\columnwidth]{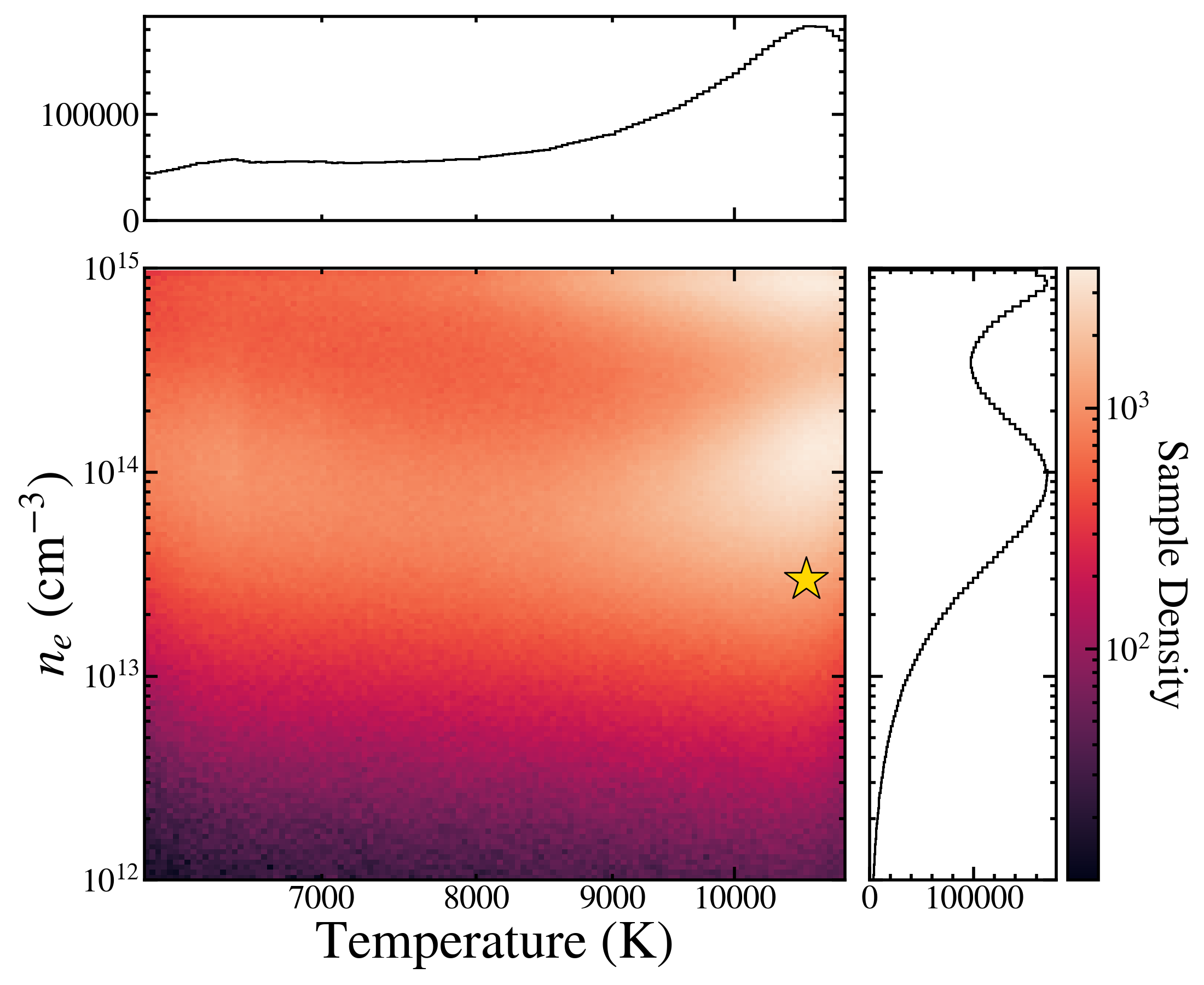}
    \caption{Joint constraints on hydrogen slab model density and temperature from the MCMC posterior sampling. The marginalized temperature and density distributions are shown in the top and right panels, respectively. The best-fit model (as determined from the MAP value) is plotted as a yellow star.} 
    \label{fig:post}
\end{figure} 
\endgroup

\cite{Herczeg2008} applied these models to spectra of young, very low mass stars and brown dwarfs and demonstrated sensitivity to accretion rates down to $\sim\!10^{-12}$ M$_\odot$ yr$^{-1}$. Many other studies have adopted the same approach to model accretion onto brown dwarfs; for example, \cite{Alcala2014} analyzed 36 stellar and substellar YSOs in Lupus star-forming region and measured typical accretion rates of 2 $\times$ 10$^{-12}$ M$_\odot$ yr$^{-1}$ to 4 $\times$ 10$^{-8}$ M$_\odot$ yr$^{-1}$. The same framework has since been used to estimate accretion rates down to planetary host masses. Slab-model fits to the UV continuum were used to estimate accretion rates of $\sim\!10^{-13}$ to $\sim\!10^{-8}$ M$_\odot$ yr$^{-1}$ for giant planets ($M_* \sim2-13$ $M_\mathrm{Jup}$) such as PDS 70 b \citep{Zhou2014, Zhou2021}. 

To determine the accretion luminosity of SR 12 c, we fit slab models to photometry in four WFC3-UVIS filters, F225W, F336W, F438W, and F555W, with photospheric emission in each broad-band filter is first subtracted from the de-reddened HST photometry. Note that while the filter transmission profile for the F555W filter includes the wavelength of the H$\alpha$ line (6563 \AA), the contribution of the line emission to the estimated continuum flux in the F555W filter is $\sim$8\%. Because the fractional uncertainty on the F555W photometry is 28\%, and therefore subsumes any significant change in continuum flux due to line emission, we do not adjust the observed flux in F555W to avoid the slight bias introduced by H$\alpha$. The H$\alpha$ filter (F656N) was not included in the fit as we seek to model the continuum excess rather than line emission. First, we generate synthetic photometry through the four bandpasses for each slab model using the {\fontfamily{pcr}\selectfont species} package \citep{Stolker2020}, resulting in a set of four flux densities per model. There are 3,300 models, each at a discrete temperature and density pairing on a grid of 11 temperatures (spanning 6000 to 11000 K) and 300 densities (spanning 1.00$\times 10^{12}$ to $\sim$9.77$\times 10^{14}$ cm$^{-3}$). These ranges encompass the diversity of inferred properties of accreting brown dwarfs and giant planets in previous studies (e.g., \citealt{Herczeg2008, Zhou2014, Alcala2014, Alcala2017, Zhou2021, Zhou2023}). 

Model parameters and associated uncertainties are determined with a custom Markov Chain Monte Carlo (MCMC) Metropolis-Hastings sampling algorithm following \cite{Metropolis} and \cite{Hastings}. To identify initialization values at which to launch the MCMC walker, we compute a $\chi ^2$ value for each model, then adopt the model $T$ and $n_e$ that yield the lowest $\chi ^2$ value. The grid of models is then interpolated to inter-grid positions for finer sampling with MCMC. Our approach to generate models at any arbitrary inter-step spacing is to linearly interpolate the grid using the {\fontfamily{pcr}\selectfont RegularGridInterpolator} function from the {\fontfamily{pcr}\selectfont scipy} package. We adopt uniform priors for both parameters (linear for $T$ and log-uniform for $n_e$), bounded by the highest and lowest temperatures and densities in the grid. The native grid spacing is $\Delta T = 500~\mathrm{K}$ and $\Delta \log_{10} n_e = 0.01~\mathrm{dex}$. Each model is first scaled to the data using the scaling constant that minimizes the $\chi ^2$ value for that particular model \citep{Cushing2008}:

\begin{equation}\label{eq:scaling}
    \hat c = \frac{\sum^4_{i=1}(\frac{f_{\lambda, i}m_i}{\sigma_i^2})}{\sum^4_{i=1}(\frac{m_i}{\sigma_i})^2}        ,
\end{equation}
where $f_{\lambda, i}$ is the observed photometry (corrected for extinction) minus the photosphere at wavelength $i$, $m_i$ represents the corresponding model synthetic photometry, and $\sigma_i$ represents the uncertainty associated with the measured flux density in each filter. 

The parameter posteriors are shown in Figure~\ref{fig:post}. The Gelman-Rubin statistic \citep{GelmanRubin1992} is monitored to ensure convergence ($\hat R \lesssim$1.02). Maximum a posteriori (MAP) values are adopted for the best-fit values, and uncertainties represent the 68\% highest-density interval. There is a preference for high temperatures, but overall this parameter is not well constrained: $T$ = 10640$^{+362}_{-2531}$ K. The number density is somewhat more precisely determined: $n_e$ = 2.17$^{+19.4}_{-2.07}\times10^{-13}$ cm$^{-3}$. The best-fit model is shown in Figure~\ref{fig:manymods}, and has a $\chi ^2$ value of 5.04. Since there are four data points and three model parameters ($T$, $n_e$, and scale factor), $\nu=1$ and the $\chi^2_{\nu}$ is equal to the $\chi ^2$. 

\begingroup 
\nolinenumbers
\begin{figure}
    \centering
    \includegraphics[width =\columnwidth]{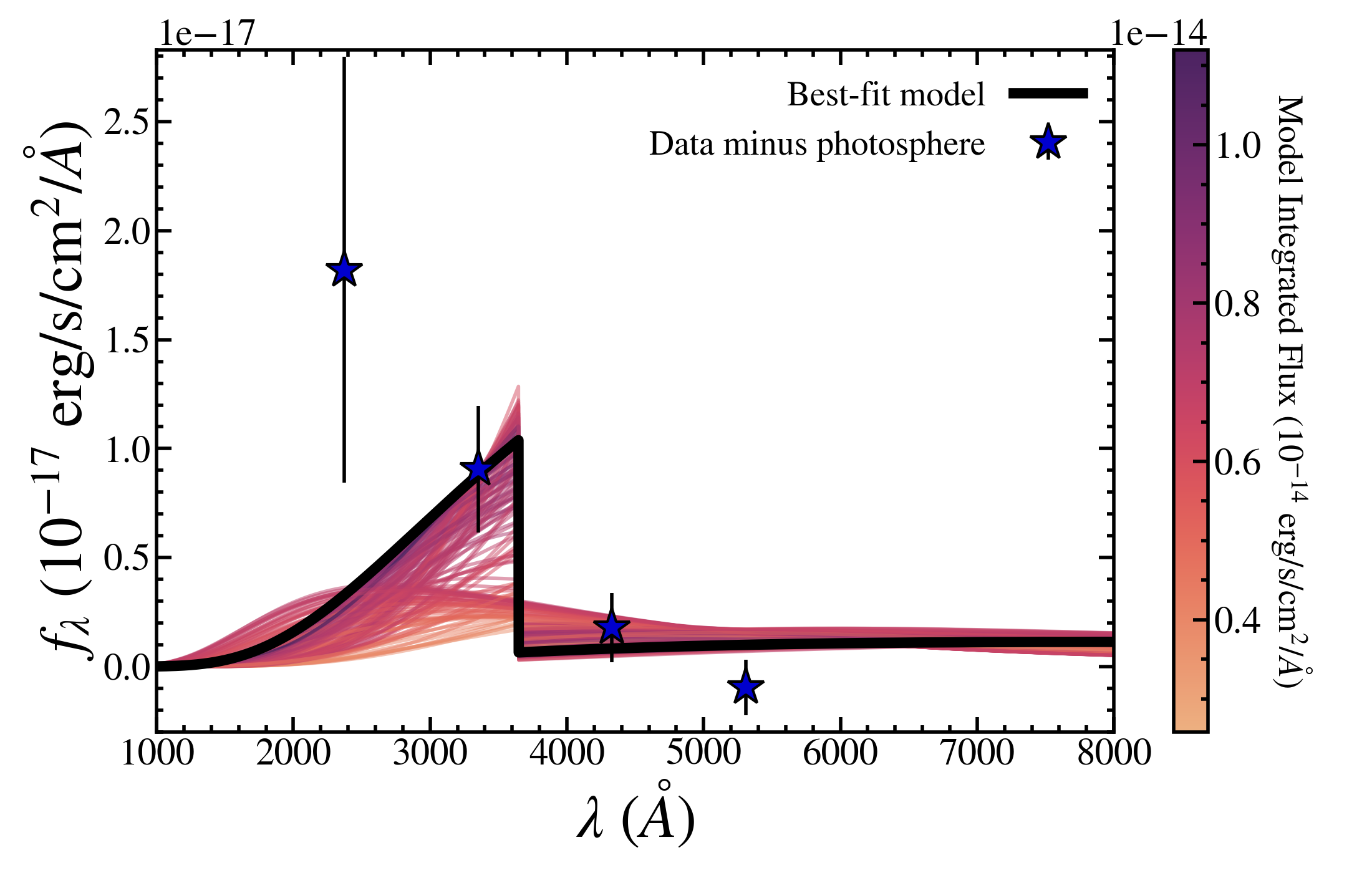}
    \caption{Best-fitting hydrogen slab model to SR 12 c (black) after removing the photosphere contribution, along with 150 random draws from the $T$ and $n_e$ posterior chains. Two qualitatively different solutions are evident: a high-temperature ($>$10000 K) family of solutions that results in a strong Balmer jump, and low-temperature ($<$8000 K) contribution with a smaller jump and a soft peak near 2000 \AA. Models are colored by total integrated flux to display the variety of Balmer jump shapes in the grid.} 
    \label{fig:manymods}
\end{figure}
\endgroup

\begingroup 
\begin{deluxetable*}{lcccc}[ht!]
\renewcommand\arraystretch{0.9}
\tabletypesize{\small}
\setlength{\tabcolsep}{0.1cm}
\tablewidth{0pt}
\tablecolumns{4}
\tablecaption{Photometry of SR 12 c \label{tab:sedphot}}
\tablehead{
\colhead{Filter} & \colhead{Central Wavelength} & \colhead{Observed Flux Density} & \colhead{De-reddened Flux Density$^a$} & Reference \\
\colhead{} & \colhead{($\mu$m)} & \colhead{($10^{-17}$ erg s$^{-1}$ cm$^{-2}$ \AA$^{-1}$)} & \colhead{($10^{-17}$ erg s$^{-1}$ cm$^{-2}$ \AA$^{-1}$)}}
\startdata
WFC3/F225W & 0.237 & 0.14 $\pm$ 0.04 & 1.8 $\pm$ 0.9 & This work\\
WFC3/F336W & 0.336 & 0.32 $\pm$ 0.03 & 0.94 $\pm$ 0.29 & This work\\
WFC3/F438W & 0.433 & 0.18 $\pm$ 0.04 & 0.44 $\pm$ 0.16 & This work\\
WFC3/F555W & 0.531 & 0.6 $\pm$ 0.2 & 0.45 $\pm$ 0.13 & This work\\
WFC3/F656N & 0.656 & 18.0 $\pm$ 0.6 & 32 $\pm$ 4 & This work\\
PAN-STARRS1/$z$ & 0.870 & 3.6 $\pm$ 0.1 & 5.2 $\pm$ 0.5 & \cite{Chambers2016} \\
PAN-STARRS1/$y$ & 0.960 & 8.8 $\pm$ 0.2 & 12.1 $\pm$ 0.8 & \cite{Chambers2016} \\
NICMOS/F110W & 1.120 & 15.0 $\pm$ 0.4 & 17.7 $\pm$ 0.9 & \cite{Allen2002} \\
2MASS/$J$ & 1.235 & 13 $\pm$ 1.0 & 15.6 $\pm$ 1.3 & \cite{Cutri2003} \\
IRSF/$J$ & 1.241 & 13.0 $\pm$ 0.4 & 15.7 $\pm$ 0.7 & \cite{Kuzuhara2011} \\
NICMOS/F160W & 1.604 & 10.0 $\pm$ 0.3 & 11.6 $\pm$ 0.4 & \cite{Allen2002} \\
IRSF/$H$ & 1.618 & 10.0 $\pm$ 0.3 & 11.4 $\pm$ 0.4 & \cite{Kuzuhara2011} \\
IRCS/$H$ & 1.635 & 8.5 $\pm$ 1.0 & 9.7 $\pm$ 1.2 & \cite{Kuzuhara2011} \\
2MASS/$H$ & 1.662 & 11 $\pm$ 1.0 & 11.9 $\pm$ 1.2 & \cite{Cutri2003} \\
IRSF/$K_s$ & 2.132 & 6.5 $\pm$ 0.2 & 7.1 $\pm$ 0.2 & \cite{Kuzuhara2011} \\
2MASS/$K_s$ & 2.159 & 7.4 $\pm$ 0.5 & 8.1 $\pm$ 0.5 & \cite{Cutri2003} \\
IRAC/3.6 $\mu$m & 3.55 & 1.7 $\pm$ 0.2 & 1.8 $\pm$ 0.2 & \cite{Martinez2022} \\
IRCS/$L'$ & 3.77 & 3.1 $\pm$ 0.2 & 3.2 $\pm$ 0.2 & \cite{Kuzuhara2011} \\
IRAC/4.5 $\mu$m & 4.49 & 1.40 $\pm$ 0.08 & 1.47 $\pm$ 0.09 & \cite{Martinez2022} \\
IRAC/5.8 $\mu$m & 5.73 & 0.60 $\pm$ 0.08 & 0.63 $\pm$ 0.09 & \cite{Martinez2022} \\
MIRI/F560W & 5.64 & 0.672 $\pm$ 0.007 & 0.683 $\pm$ 0.007 & Wu et al. (submitted)\\
MIRI/F770W & 7.64 & 0.358 $\pm$ 0.003 & 0.364 $\pm$ 0.004 & Wu et al. (submitted)\\
IRAC/8.0 $\mu$m & 7.87 & 0.40 $\pm$ 0.05 & 0.43 $\pm$ 0.05 & \cite{Martinez2022} \\
MIRI/F1000W & 9.95 & 0.222 $\pm$ 0.002 & 0.235 $\pm$ 0.003 & Wu et al. (submitted)\\
MIRI/F1280W & 12.81 & 0.122 $\pm$ 0.002 & 0.124 $\pm$ 0.002 & Wu et al. (submitted)\\
MIRI/F1500W & 15.06 & 0.0889 $\pm$ 0.0024 & 0.0904 $\pm$ 0.0025 & Wu et al. (submitted)\\
MIRI/F1800W & 17.98 & 0.0621 $\pm$ 0.0028 & 0.0635 $\pm$ 0.0029 & Wu et al. (submitted)\\
MIRI/F2100W & 20.80 & 0.0523 $\pm$ 0.0035 & 0.0533 $\pm$ 0.0036 & Wu et al. (submitted)\\
ALMA/Band 7 & 880 & (5 $\pm$ 0.5) $\times$ $10^{-6}$ & (5 $\pm$ 0.5) $\times$ $10^{-6}$ & \cite{Wu2022}\\
\enddata
\tablecomments{$^a$Observed photometry de-reddened by $A_V=0.85$ $\pm$ 0.17 mag.}
\end{deluxetable*} 
\endgroup

\section{Discussion} \label{sec:disc}

\subsection{Accretion onto SR 12 c}\label{sec:accdisc} 
Accretion luminosity emerges as the combination of hot continuum emission and line emission. The continuum dominates in the UV, and line emission arises in the UV (Lyman series), optical (Balmer series), and infrared (Paschen, Brackett, Pfund, and Humphreys series). Both continuum excess and H$\alpha$ accretion are clearly evident in the HST photometry of SR 12 c. Here, we describe our procedure for estimating the total accretion luminosity as the sum of continuum-plus-line luminosity. To measure the continuum contribution to the accretion flux, we integrate under each slab model in our MCMC posterior distribution from 100 to 100,000 \AA. Using the distance to the system (139 $\pm$ 5 pc), we converted this continuum flux to a continuum luminosity and propagate the uncertainty in a Monte Carlo fashion. Hereinafter we refer to this as the ``slab accretion luminosity" or $L_\mathrm{slab}$ to distinguish it from the additional contribution to accretion luminosity from emission lines. The resulting slab accretion luminosity is 1.32$\pm$ $0.19\times 10^{-5} L_{\odot}$, or log ($L_\mathrm{slab}$/$L_\mathrm{\odot}$) = --4.89 $\pm$ 0.08 dex. 

The uncertainty on $L_\mathrm{slab}$ remains small considering the scale of the photometric uncertainties; however, these correspond to measurement uncertainties. Recent studies (e.g. \citealt{Marleau2022}) suggest there may be other effects such as dust extinction in the local environment of the planet that could contribute substantial systematic uncertainty at a scale much larger than the measurement uncertainty. Though local extinction is challenging to measure, it could significantly increase the intrinsic accretion luminosity, and in turn the formal uncertainty on $L_\mathrm{slab}$. An inclined disk can also impact the measured value of $L_\mathrm{slab}$, although the orientation of SR 12 c's disk is not well constrained. Because SEDs of young stars exhibit strong degeneracies---particularly between inclination and disk/envelope structure---inclination estimates from SED fitting alone are generally unreliable (e.g., \citealt{Robitaille2007}). More robust constraints would require spatially resolved observations (e.g., scattered‑light imaging or ALMA kinematics).

The line luminosity from H$\alpha$ is calculated by synthesizing the F656N-band monochromatic flux density from the photosphere model and subtracting this from the measured F656N flux density of SR 12 c after correcting for extinction. We then integrate over the bandpass using the FWHM of the F656N filter, 17.5 \AA, and de-redden the result to obtain the H$\alpha$ line flux, 5.41 $\pm$ $0.68\times 10^{-15}$ erg s$^{-1}$ cm$^{-2}$. We convert this to a line luminosity using the adopted distance of 139 $\pm$ 5 pc, resulting in a line accretion luminosity of 3.29 $\pm$ $0.42 \times 10^{-6} L_\mathrm{\odot}$, or log ($L_\mathrm{line}$/$L_\mathrm{\odot}$) = --5.48 $\pm$ 0.06 dex. The implied fraction of accretion luminosity emitted in H$\alpha$ relative to the slab (accretion shock) is 0.25 $\pm$ 0.06. 

The total $L_\mathrm{acc}$ is the sum of the slab accretion luminosity and the accretion luminosity from emission lines. Other hydrogen emission lines (most notably Ly$\alpha$ \citealt{Aoyama2018}) should be considered when computing the total accretion luminosity; however, we generally do not have access to these. Inferring the total line emission from H$\alpha$ alone is model-dependent and relies on assumptions of the emitting surface and its geometry \citep{Aoyama2018, Aoyama2020}. For these purposes, we follow \cite{Zhou2014} and \cite{Zhou2021} and use the H$\alpha$ line luminosity alone, although it is clear that this will significantly underestimate the total contribution from emission lines. Nevertheless, following this approach we obtain a final $L_\mathrm{acc}$ measurement of 1.65$\pm$ $0.19 \times 10^{-5} L_{\odot}$ or log ($L_\mathrm{acc}$/$L_\mathrm{\odot}$) = --4.78 $\pm$ 0.06 dex.  

We compare this value with what we would have arrived at if we had only detected an H$\alpha$ emission line, which can tentatively serve as a proxy for accretion rate. Several attempts have been made to connect the emission strength from prominent emission lines like H$\alpha$ to the bolometric accretion luminosity. Using the relationship from \cite{Alcala2017}, our H$\alpha$ line luminosity corresponds to log $L_\mathrm{slab}$ = --4.46 $\pm$ 0.34 dex, or 3.5$^{+4.1}_{-1.9} \times 10^{-5}$ L$_\odot$. This agrees with our inferred value for log $L_\mathrm{slab}$ of --4.78 $\pm$ 0.06 dex (1.32 $\pm$ $0.19\times 10^{-5} L_{\odot}$) at the 0.9 $\sigma$ level, indicating the relationship from \cite{Alcala2017}, calculated based on young stellar objects in the Lupus cloud ($\sim$3 Myr, \citealt{Alcala2017}), yields reasonable estimates for this planetary-mass companion in the comparably aged $\rho$ Oph region \citep{Wilking2008, EsplinLuhman2020}. 

The mass accretion rate, $\dot M$, can be estimated following \cite{Gullbring1998}:

\begin{equation}\label{eq:massacc}
    \dot M \approx \frac{1.25R_*L_\mathrm{acc}}{GM_*}
\end{equation}

\noindent Here, the factor of 1.25 implies that there is an inner accretion disk radius, or a magnetospheric truncation radius of 5 $R_*$, which in the context of magnetospheric accretion is attributed to a magnetospheric truncation radius ($R_\mathrm{in}$). The value of $R_\mathrm{in}$ is uncertain as it relies on details such as accretion flow coupling to the planet's magnetic field, but it must inherently be lower than the co-rotation radius between the disk and the planet in order to allow accretion onto the planet. A commonly adopted co-rotation radius is 5 $R_*$, simplifying the prefactor $\big(1- R_*/R_\mathrm{in}\big)^{-1}$ to 1.25. Recent studies (e.g., \citealt{Pittman2025}) have found an $R_\mathrm{in}$ of 2.5$R_*$ may be more appropriate for stars, but considering the uncertainty surrounding this value in the planetary-mass regime, we adopt the conventional 5 $R_*$.

Using the effective temperature of 2600 $\pm$ 100 K and the bolometric luminosity of log($L/L_{\odot}$)= $-2.82 \pm 0.10$ dex (Wu et al., submitted), the Stefan-Boltzmann law yields a radius of 1.9 $\pm$ 0.3 $R_\mathrm{Jup}$. This radius is $\sim$0.5 $R_\mathrm{Jup}$ lower than the BHAC15 \citep{Baraffe2015} prediction, perhaps hinting that the system is slightly older than the median. Combining the radius with the planet's assumed mass from evolutionary models \citep{Baraffe2015}, 16 $\pm$ 2 $M_\mathrm{Jup}$, this implies a mass accretion rate of 8 $\pm$ $2\times 10^{-12}$ M$_{\odot}$ yr$^{-1}$, or log $\dot M$/M$_{\odot}$ = --11.1 $\pm$ 0.1 dex. 

At the current rate, it would take 1.8 Gyr for SR 12 c to accrete its 16 $M_\mathrm{Jup}$---far longer than the planet's short 3-4 Myr lifetime. Consequently, over the next Myr, SR 12 c will only gain $\sim$0.06\% of its current mass. This means that SR 12 c must be at the end of its accretion phase and must have had a much higher accretion rate in the past. This result implies a decaying mass accretion process, and is consistent with the expected giant planet formation timescales of a few Myr \citep{Alibert2004}. 

\subsubsection{Accretion Variability}\label{sec:var}
Recent observations of young, accreting planets like PDS 70 b and c \citep{Zhou2025, Close2025a} and the candidate protoplanet AB Aur b \citep{Currie2022, Zhou2023, Bowler2025} show clear evidence that H$\alpha$ varies on timescales of days to years. Variability in planetary-mass accretion rates indicates a non-static accretion process, and implies some degree of stochastic mass transfer from the circumplanetary accretion disk. This is not surprising, given that typical protostellar accretion is known to be highly variable on timescales from hours to years (e.g., \citealt{Hartmann2016}).

The HST/WFC3 observations were taken with an identical setup in all five filters in two epochs on UT 2021 February 28 and UT 2021 March 26. Here, we examine potential variability spanning this $\sim$month-long baseline. Comparing the observed flux densities from Epoch 1 to Epoch 2, we do not find signs of significant ($>$3$\sigma$) variability. Figure~\ref{fig:variability} shows a depiction of the percent change in flux between HST observations. Given the non-detection of variability in all UV and optical filters, we have co-added these epochs and adopted the resulting photometry for this study (see Section~\ref{sec:hst}). Individual and averaged measurements for all filters are reported in Table~\ref{tab:hstphotlog}.

\begingroup 
\nolinenumbers
\begin{figure}
    \centering
    \includegraphics[width =\columnwidth]{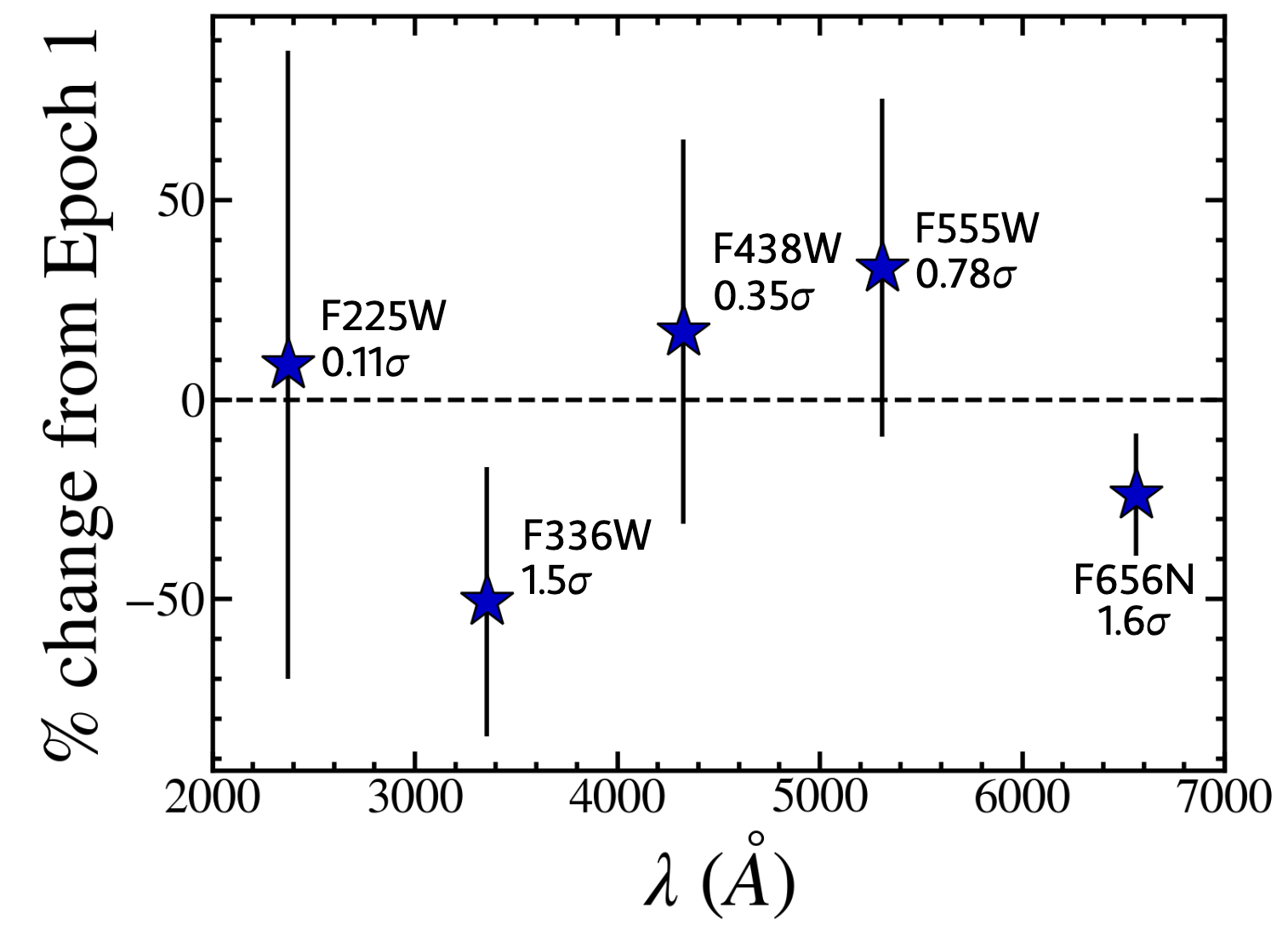}
    \caption{The percent change in flux of SR 12 c from Epoch 1 to Epoch 2 of the HST observations, plotted as a function of the central wavelength of each filter. There is no evidence for significant variability in any of the filters over the month-long baseline.} 
    \label{fig:variability}
\end{figure}
\endgroup
 
We also examine variability on the timescale of years by comparing our results with previous work. \cite{SantamariaMiranda2018} reported the H$\alpha$ line strength of SR 12 c from an X-shooter spectrum acquired on UT 2016 May 2. They found a line flux of 7.69 $\pm$ $2.39\times 10^{-14}$ erg s$^{-1}$ cm$^{-2}$. Because this value was calculated assuming an $A_V$ of 1.24 mag, we adjusted it to align with our extinction of 0.85 mag, which yields a line flux of 5.85 $\pm$ 1.82 $\times 10^{-14}$ erg s$^{-1}$ cm$^{-2}$. Our inferred line flux for SR 12 c from the co-added image is 91 $\pm$ 3\% lower at 5.41 $\pm$ $0.68\times 10^{-15}$ erg s$^{-1}$ cm$^{-2}$. It is clear that the accretion luminosity emerging from the H$\alpha$ line is considerably variable over long timescales. These results for SR 12 c are in line with the conclusions presented in \cite{Demars2023}: emission line variability for PMCs tends to be mild or moderate on shorter timescales ($<$50\% on month-long timescales), and more significant on a longer timescale ($\sim$1000\% over a multi-decade baseline), similar to the patterns noted for classical T Tauri stars. 

\begingroup 
\nolinenumbers
\begin{deluxetable*}{lccccccc}[ht]
\renewcommand\arraystretch{0.9}
\tabletypesize{\small}
\setlength{ \tabcolsep } {.1cm}
\tablewidth{0pt}
\tablecolumns{8}
\tablecaption{Accretion properties of young PMCs as measured with UV continuum excess \label{tab:laccs}}
\tablehead{
 \colhead{Object} & \colhead{Mass} & \colhead{log $L_\mathrm{H\alpha}/L_{\odot}$} & \colhead{log $L_\mathrm{slab}/L_{\odot}$} & \colhead{log $L_\mathrm{acc}/L_{\odot}$} & \colhead{$L_\mathrm{H\alpha}$/$L_\mathrm{slab}$} & \colhead{log $\dot M$} & \colhead{Ref.} \\
 \colhead{} & \colhead{($M_\mathrm{Jup}$)} & \colhead{[dex]} & \colhead{[dex]} & \colhead{[dex]} & \colhead{} & \colhead{($M_\odot$ yr$^{-1}$)} & \colhead{}
 }
\startdata
SR 12 c & 16 $\pm$ 2 & --5.48 $\pm$ 0.6 & --4.89 $\pm$ 0.08 & --4.78 $\pm$ 0.06 & 0.25 $\pm$ 0.06 & --11.1 $\pm$ 0.1 & This work\\
GQ Lup B & 33 $\pm$ 10 & --4.69 & --2.91 & --2.90 & 0.017 & --9.30 & 1, 4\\
GSC 06214-00210 b & 14.5 $\pm$ 2.0 & --5.03 & --4.65 & --4.60 & 0.42 & --10.9 & 2, 4\\
DH Tau b & 12 $\pm$ 4 & --6.19 & --5.40 & --5.30 & 0.16 & --11.4 & 1, 4\\
PDS 70 b & 3.2$^{+8.4}_{-2.1}$ & --6.19 $\pm$ 0.06 & --5.93 $\pm$ 0.08 & --5.75 $\pm$ 0.05 & 0.54 $\pm$ 0.12 & --10.9 $\pm$ 0.06 & 3, 5\\
\enddata

\tablecomments{(1) Mass from \cite{Xuan2024}.}

\tablecomments{(2) Mass from \cite{Pearce2019}.}

\tablecomments{(3) Mass from \cite{Wang2021}.}

\tablecomments{(4) Accretion properties from \cite{Zhou2014}.}

\tablecomments{(5) Accretion properties from \cite{Zhou2021}.}
\end{deluxetable*}
\endgroup

\begingroup 
\nolinenumbers
\begin{figure}
    \centering
    \includegraphics[width =\columnwidth]{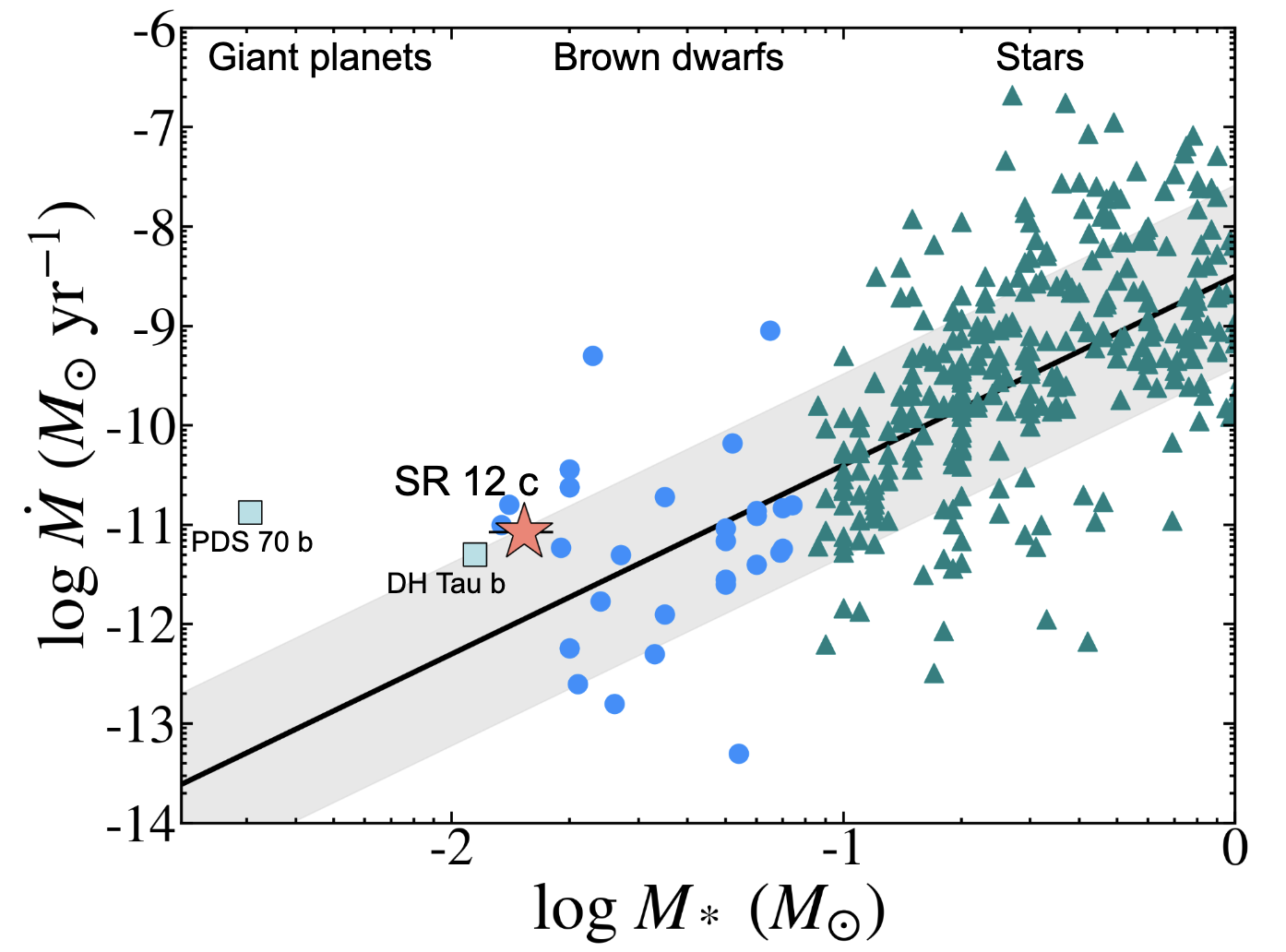}
    \caption{Log $\dot M$ versus log $M_*$ for accreting stellar and substellar objects as measured by UV continuum excess from the CASPAR database \citep{Betti2023}, \cite{Zhou2014, Zhou2021}, and this work. Stars are shown as teal triangles, brown dwarfs are blue circles, and planetary-mass objects are in blue squares. SR 12 c is shown as a coral star. The best-fit relation from \cite{Betti2023}, $\dot M \sim M^{2.16}$, is in black and the 1$\sigma$ upper limit for this relation is shaded in gray.} 
    \label{fig:casp}
\end{figure}
\endgroup

\subsection{Comparison to Other Accreting Substellar Objects with Measured Continuum Excess}\label{sec:acccom}
There are a growing number of accreting bound and free-floating planetary-mass objects with accretion rate estimates (cataloged in the CASPAR database, \citealt{CASPAR}), but most of these are based solely on H$\alpha$ emission. Here we summarize results for those with accretion rates determined from UV excess using observations from HST/WFC3. 

\cite{Zhou2014} reported $L_\mathrm{acc}$ and $\dot M$ values for three young substellar objects, two of which are similar to SR 12 c in mass ($\approx$12-15 $M_\mathrm{Jup}$): GSC 06214-00210 b \citep{Ireland2011, Zhou2014, Demars2023, Xuan2024} and DH Tau b \citep{Itoh2005, Xuan2024}. The third object, GQ Lup B \citep{Neuhauser2005}, has a somewhat higher mass estimate of 10--40 M$_\mathrm{Jup}$ \citep{Marois2007, Seifahrt2007, Cugno2024, Xuan2024}. \cite{Zhou2021} also reported accretion luminosity and mass accretion rate measurements for the protoplanet PDS 70 b ($\sim$3 $M_\mathrm{Jup}$, \citealt{Wang2021}). 

In Table~\ref{tab:laccs}, we summarize previous measurements of log $L_\mathrm{H\alpha}$, log $L_\mathrm{slab}$, log $L_\mathrm{acc}$, $L_\mathrm{H\alpha}$/$L_\mathrm{slab}$, and $\dot M$ for these objects along with our measurements for SR 12 c. Among these five objects, SR 12 c falls in the middle of the sample in terms of accretion luminosity. It has an accretion luminosity most similar to GSC 06214-00210 b, which is slightly older at 5--10 Myr \citep{Bowler2015}. SR 12 c has a $L_\mathrm{H\alpha}$/$L_\mathrm{slab}$ fraction most similar to DH Tau b, which is similarly young ($\approx$3 Myr, \citealt{Itoh2005}). GQ Lup B has the highest accretion luminosity, and corresponds to the highest-mass companion in the sample. Interestingly, it has the lowest $L_\mathrm{H\alpha}$/$L_\mathrm{acc}$ fraction by an order of magnitude. SR 12 c's mass accretion rate is most consistent with that of PDS 70 b and GSC 06214-00210 b. Note, however, that direct comparisons among these objects is complicated by many factors---companion mass, system age, viewing geometry, and accretion variability. 

In Figure~\ref{fig:casp}, planetary-mass companions SR 12 c, PDS 70 b, and DH Tau b fall above the well-studied $\dot M-M_*$ relation \citep{Calvet2004, Natta2004, Muzerolle2005, Herczeg2008, Hartmann2016, Betti2023}. This may be due to an emerging break in the $\dot M-M_*$ relation at planetary masses, which has been suggested to serve as an indication of formation through disk instability \citep{StamWhit2009, StamHerc2015}. While the current accretion rate does not directly reflect the epoch when SR 12 c assembled most of its mass, companions formed through fragmentation may retain massive disks and continue accreting for several Myr. Alternatively, bound planetary-mass companions may present higher accretion rates than their higher-mass free-floating brown dwarf counterparts due to the reservoir of available gas in the circumstellar disk to feed accretion. Or perhaps it is simply the result of a selection bias favoring more heavily accreting objects. A larger sample of accreting planets is needed to distinguish these scenarios. 

A subset of PMCs hosting CPDs such as GSC 6214–210 b, Delorme 1 (AB) b, and now SR 12 c, illustrate systems in which the host stars show no detectable circumstellar or circumbinary disk, while their wide planetary‑mass companions retain actively accreting disks \citep{Bowler2015, Eriksson2020}. This indicates differential disk‑depletion timescales between the host and companion, reflecting intrinsic differences in disk evolution, or it may be a result of previous dynamical interactions.

\section{Conclusion} \label{sec:conc}
In this work, we present newly acquired HST/WFC3 imaging for the wide-orbit, accreting planetary-mass companion SR 12 c which hosts a circumplanetary disk \citep{Martinez2022, Wu2022}. We report UV-through-red optical photometry for SR 12 c and combine it with previous observations and new JWST observations to construct an SED with the most-complete wavelength coverage for a young, accreting giant planet to date. We also report an updated estimate of the extinction to the system of $A_V$ = 0.85 $\pm$ 0.17 mag based on photometry of SR 12 AB. We model the UV through mid-IR SED with three contributions: a hydrogen slab model that dominates in the UV, a photosphere model, and a two-temperature blackbody disk model described in detail in Wu et al., (submitted). 

We infer an accretion luminosity of 1.65$\pm$ $0.19\times 10^{-5} L_{\odot}$ and mass accretion rate of 8 $\pm$ $2\times 10^{-12}$ M$_{\odot}$ yr$^{-1}$ by fitting hydrogen slab models to the UV photometry across the Balmer jump. We find this derived accretion luminosity to be in agreement with the inferred accretion luminosity derived from the H$\alpha$ flux using the relations in \cite{Alcala2017}, affirming that the connection between line luminosity and continuum excess persists down to planetary masses. We reinforce the general conclusions presented in \cite{Demars2023} for other planetary-mass companions, finding the H$\alpha$ emission of SR 12 c is not significantly variable over a short baseline of a month, but varies by 90\% over a longer baseline of five years.

When taking the accretion properties of SR 12 c in conjunction with results from the 5.6--21 $\mu$m JWST/MIRI photometry, a picture emerges of a young planet that has nearly finished accreting gas from its circumplanetary disk, which itself has undergone a level of grain growth and dust settling that is indicative of a later evolutionary stage (Wu et al., submitted.) At the present rate of accretion, SR 12 c will only increase in mass by $\sim$0.06\% over the next Myr; this seems to be the final stage of growth for this object. Expanding the sample of wide orbit and free-floating giant planets with measurements of accretion in the UV and CPD structure in the mid-IR and sub-mm wavelengths is key to informing the timescale and mechanism by which protoplanets accrete mass, constraining the timescale of giant planet and potentially exomoon formation.

\section{Acknowledgments}
C.O.F. is thankful to Lillian Jiang for her assistance in employing the hydrogen slab models. B.P.B. acknowledges support from the National Science Foundation grant AST-1909209, NASA Exoplanet Research Program grant 20-XRP20$\_$2-0119, and the Alfred P. Sloan Foundation. This research is based on observations made with the NASA/ESA Hubble Space Telescope obtained from the Space Telescope Science Institute. The data were obtained from the Mikulski Archive for Space Telescopes at the Space Telescope Science Institute, which is operated by the Association of Universities for Research in Astronomy, Inc., under NASA contract NAS 5–26555 for HST. These observations are associated with program GO 16302. This work is also based on observations made with the NASA/ESA/CSA James Webb Space Telescope. The data were obtained from the Mikulski Archive for Space Telescopes at the Space Telescope Science Institute, which is operated by the Association of Universities for Research in Astronomy, Inc., under NASA contract NAS 5-03127 for JWST. These observations are associated with program GO 2311.

\bibliographystyle{aasjournal}
\bibliography{references.bib}

\end{document}